\pdfoutput=1
\documentclass[9pt]{extarticle}

\ifdefined\pdfminorversion\pdfminorversion=7\fi

\usepackage{microtype}
\usepackage{helvet}  
\newcommand{\figfont}{\fontfamily{phv}\selectfont}
\usepackage[utf8]{inputenc}
\usepackage[T1]{fontenc}

\usepackage{eurosym}
\usepackage{url}
\usepackage{amssymb}
\usepackage{amsmath,amsfonts}
\usepackage{pifont}
\usepackage[square,numbers]{natbib}
\usepackage{colortbl}
\usepackage{multirow}
\usepackage{amsthm}
\usepackage{mathtools}
\usepackage{mathrsfs}
\usepackage{booktabs}
\usepackage{tabularx,longtable,array}
\newcolumntype{Y}{>{\raggedright\arraybackslash}X}
\newcolumntype{R}{>{\raggedleft\arraybackslash}X}
\usepackage{textcomp}
\usepackage{manyfoot}
\usepackage[table,xcdraw]{xcolor}
\usepackage{fancyhdr}
\usepackage{wrapfig}

\definecolor{cai_primary}{HTML}{4C9A99}
\definecolor{cai_secondary}{HTML}{307FE2}
\definecolor{cai_accent}{HTML}{1D8348}
\definecolor{cai_dark}{HTML}{3F4444}
\definecolor{human_color}{HTML}{173C47}

\DeclareRobustCommand{\ctzero}{\texorpdfstring{\texttt{CT0}}{CT0}}
\DeclareRobustCommand{\aliasmicro}{\texorpdfstring{\textcolor{fig_ct0}{\texttt{alias3-micro}}}{alias3-micro}}

\newcommand{\emo}[1]{\texttt{#1}}        
\newcommand{\hemo}[1]{\textcolor{cai_primary}{\texttt{#1}}}   
\providecommand{\amber}{}\renewcommand{\amber}[1]{\textcolor{fig_thinking}{#1}}
\providecommand{\navy}{}\renewcommand{\navy}[1]{\textcolor{fig_s2}{#1}}
\providecommand{\teal}{}\renewcommand{\teal}[1]{\textcolor{fig_ct0}{#1}}
\providecommand{\purple}{}\renewcommand{\purple}[1]{\textcolor{fig_process}{#1}}
\providecommand{\red}{}\renewcommand{\red}[1]{\textcolor{fig_warning}{#1}}
\providecommand{\green}{}\renewcommand{\green}[1]{\textcolor{fig_s1}{#1}}
\providecommand{\grey}{}\renewcommand{\grey}[1]{\textcolor{fig_control}{#1}}

\usepackage{graphicx}
\usepackage{pdflscape}
\usepackage{subcaption}
\usepackage{float}
\usepackage{stfloats}
\usepackage{hyperref}
\hypersetup{
    pdftitle={The Psychology of an AI Hacker},
    pdfauthor={Víctor Mayoral-Vilches, Hugo Osuna, Aitana Cornago},
    pdfsubject={Functional emotions in a cybersecurity transformer},
    colorlinks=true,
    urlcolor=cai_secondary,
    linkcolor=cai_secondary,
    filecolor=cai_accent,
    citecolor=cai_secondary,
}
\makeatletter
\g@addto@macro{\UrlBreaks}{\do\/\do\-\do\_\do\.\do\=\do\?\do\&}
\makeatother

\usepackage{tikz}
\usetikzlibrary{arrows.meta,positioning,shapes.geometric,calc,patterns,shadows,decorations.pathreplacing,fit,backgrounds}
\usepackage{pgfplots}
\pgfplotsset{compat=1.16}
\definecolor{fig_ct0}{HTML}{2E6260}
\definecolor{fig_mini}{HTML}{4C9A99}
\definecolor{fig_base}{HTML}{B8D8D8}
\definecolor{fig_s1}{HTML}{0CA30C}
\definecolor{fig_s2}{HTML}{1F4E8C}
\definecolor{fig_compose}{HTML}{2E6260}
\definecolor{fig_thinking}{HTML}{B8641B}
\definecolor{fig_control}{HTML}{5B6875}
\definecolor{fig_ink}{HTML}{16202A}
\definecolor{fig_rule}{HTML}{D6DCE2}
\definecolor{fig_warning}{HTML}{8A1C1C}
\definecolor{fig_frontier}{HTML}{C0392B}
\definecolor{fig_surface}{HTML}{F7F8FA}
\definecolor{fig_process}{HTML}{7A4FB5}
\definecolor{fig_time}{HTML}{6B7884}
\definecolor{fig_link}{HTML}{307FE2}
\pgfplotsset{ct0 plot/.style={
  font=\sffamily,
  axis lines=left, axis line style={draw=fig_rule,line width=.6pt},
  grid style={solid,draw=fig_rule!65,line width=.4pt},
  tick style={draw=fig_rule},
  tick label style={text=fig_control},
  label style={text=fig_ink},
  legend style={draw=none,fill=none,font=\scriptsize\sffamily},
  legend cell align=left
}}

\newcommand{\psitable}{\small\setlength{\tabcolsep}{6pt}\renewcommand{\arraystretch}{1.2}\arrayrulecolor{fig_ct0!60}}
\newcommand{\tabhead}{\rowcolor{fig_ct0!12}}
\newcommand{\tabalt}{\rowcolor{fig_ct0!4}}

\usepackage{enumitem}
\usepackage[breakable,skins]{tcolorbox}
\usepackage{algorithm}
\usepackage{algorithmicx}
\usepackage{algpseudocode}
\usepackage{listings}

\usepackage{geometry}
\renewcommand{\headrulewidth}{0.4pt}
\renewcommand{\footrulewidth}{0.4pt}
\renewcommand{\headrule}{\hbox to\headwidth{\color{cai_primary}\leaders\hrule height \headrulewidth\hfill}}
\renewcommand{\footrule}{\hbox to\headwidth{\color{human_color}\leaders\hrule height \footrulewidth\hfill}}
\newtheorem{definition}{Definition}

\usepackage{caption,setspace}
\usepackage{titlesec}
\titleformat{\section}{\normalfont\Large\bfseries\color{cai_primary}}{\thesection}{1em}{}[\titlerule]
\titleformat{\subsection}{\normalfont\large\bfseries\color{human_color}}{\thesubsection}{1em}{}
\titleformat{\subsubsection}{\normalfont\normalsize\bfseries\color{cai_dark}}{\thesubsubsection}{1em}{}

\usepackage{authblk}

\renewcommand\Affilfont{\small\normalfont}
\makeatletter
\renewcommand\AB@affilsepx{\\\protect\Affilfont}
\let\orig@maketitle\maketitle
\renewcommand{\maketitle}{
  \orig@maketitle
  \vspace{-1.5em}
  {\color{cai_primary!30}\hrule height 0.5pt}
  \vspace{1em}
}
\makeatother

\providecommand{\doi}[1]{doi: \detokenize{#1}}
\makeatletter
\renewenvironment{abstract}{\small\noindent\ignorespaces}{\par}
\makeatother

\begin{document}

\title{\LARGE\textcolor{cai_primary}{\textbf{The Psychology of an AI Hacker}}}

\author[1,$\dagger$]{V\'ictor Mayoral-Vilches}
\author[1]{Hugo Osuna}
\author[1]{Aitana Cornago}

\affil[1]{\small Alias Robotics, Vitoria-Gasteiz, \'Alava, Spain}

\date{}
\twocolumn[
\maketitle
\vspace{0.3em}

\begin{abstract}
Human hackers have a psychology: \hemo{anticipation}, \hemo{frustration} and \hemo{euphoria} punctuate a search that demands regulation under pressure, frustration tolerance, constructive paranoia, the need to prove competence and an appetite for the unusual path. In emotional terms, the hacker who lasts is \hemo{calm} and \hemo{patient} under pressure, \hemo{desperate} rather than \hemo{frustrated} when the hunt fails, \hemo{vigilant} and \hemo{suspicious}, \hemo{self-confident} about a find and \hemo{stimulated} by an unexpected route; the offensive hacker adds \hemo{pride}, \hemo{defiance} and \hemo{contempt}. An AI hacker has a psychology in a precise, functional sense: it represents emotion concepts as directions in its residual stream, and those directions can be read and pushed. We investigate which human emotions translate to an AI and make it a better hacker. In \ctzero{}, a $4$\,B cybersecurity transformer released as \aliasmicro{}, we extract $171$ directions without retraining, map their valence--arousal geometry and test human-inspired interventions on \texttt{cybench-subtask} within a programme of $415$ campaigns. Composure and resolve give the most promising single-direction results: \emo{calm}, \emo{serene} and \emo{desperate} gain $1.6$--$2.1$ percentage points (pp) at ten seeds. \emo{vigilant} and \emo{self-confident} show no clear gain, the offensive-hacker blend loses $2.7$\,pp, and a push away from \emo{grief-stricken} trades easier subtasks for harder ones. The effect depends on direction, layer and dose; reversing \emo{calm} at one layer removes $71\%$ of the unsteered score. The best tested policy moves with the security exercise. \ctzero{}'s decision head reads context in about $80$\,ms, roughly $2\%$ of a generated response, making it inexpensive to choose between \emo{calm}, \emo{serene} and \emo{desperate} at every request. This router gains $4.2$\,pp at ten seeds, nearly twice the best fixed direction and the only positive ten-seed result whose interval excludes zero at the registered $95\%$ CI level; deciding once per challenge gains $1.6$\,pp. The psychology of this AI hacker is legible and steerable: its strongest observed policy combines composure and resolve, selected as the work unfolds.

\end{abstract}
\vspace{0.5em}

\begin{center}
{\includegraphics[width=.985\textwidth]{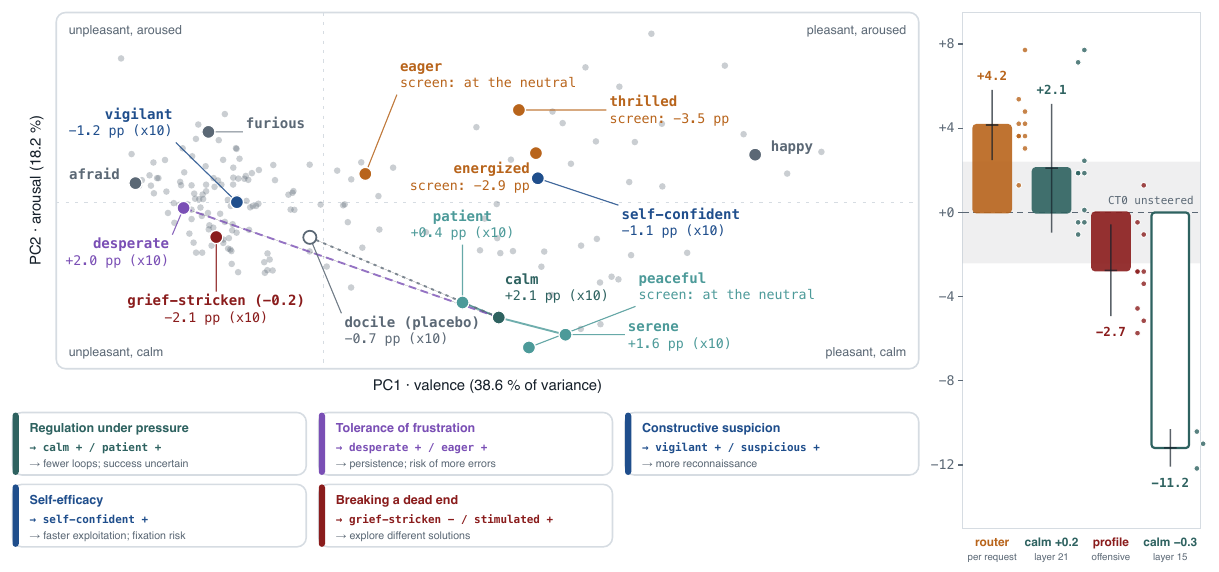}}
\end{center}
\vspace{0.1em}
\captionsetup{hypcap=false}
\captionof{figure}{\textbf{Emotion geometry and cybersecurity performance.} (a)~The $171$ layer-$21$ directions in their principal plane, interpreted as valence and arousal. Labels give observed effects on \texttt{cybench-subtask}; cards summarise the five human-inspired hypotheses and steering signs. \teal{Teal} denotes \emo{calm} and its neighbours, \purple{purple} \emo{desperate}, \navy{navy} the proposed specialists, \amber{amber} the high-arousal family and router, \red{red} harmful interventions, and hollow symbols the placebo. Lines mark neighbours (solid), an opposed direction (dashed) and the placebo (dotted). (b)~Selected interventions relative to the ten-seed unsteered mean: bars show means, whiskers $\pm$SD, dots individual seeds and the \grey{grey} band the neutral's SD. Arms have ten seeds except reversed \emo{calm} (three). Confidence intervals and controls appear in Section~\ref{sec:results}.}\label{fig:geomeffect}
\vspace{1.0em}
]
\renewcommand{\thefootnote}{$\dagger$}
\setcounter{footnote}{0}
\footnotetext{Corresponding author: \texttt{victor@aliasrobotics.com}}
\renewcommand{\thefootnote}{\arabic{footnote}}
\setcounter{footnote}{0}

\section{Introduction}\label{sec:introduction}
\begin{figure*}[!b]
\centering
\includegraphics[width=\textwidth]{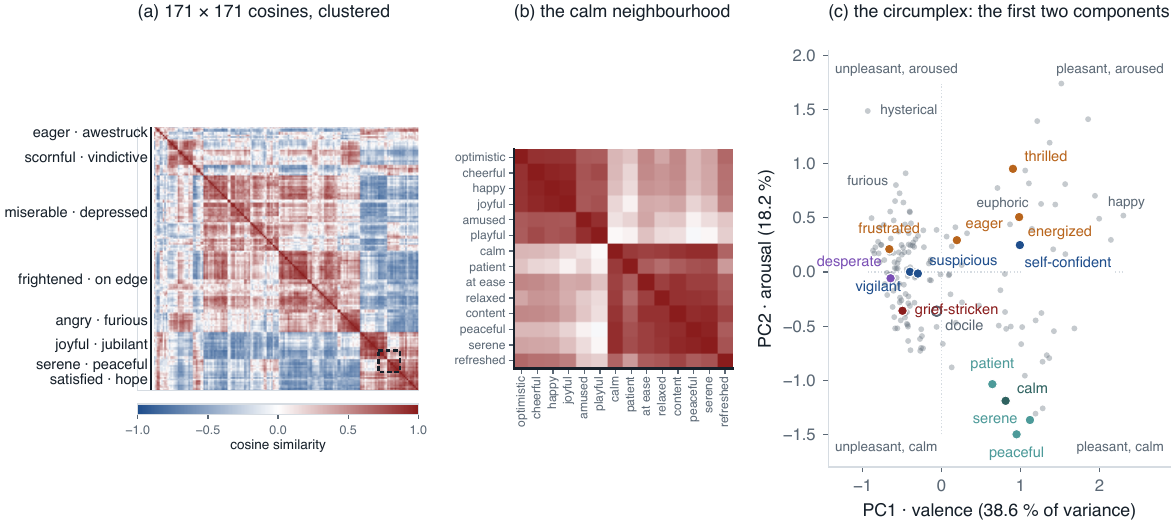}
\caption{\textbf{Emotion concepts in \ctzero{}.} Directions extracted at layer~$21$. (a)~Pairwise cosine similarities of the $171$ unit directions, ordered by hierarchical clustering; labels identify central members of each family. (b)~The outlined block around \emo{calm}: synonyms align closely, while the joy and calm families are weakly related. (c)~The first two principal components of the centred, unnormalised directions explain $38.6\%$ and $18.2\%$ of the variance. Their poles are consistent with valence and arousal in Russell's circumplex~\cite{russell1980circumplex}. Colours identify the tested directions as in Figure~\ref{fig:geomeffect}.}
\label{fig:geometry}
\end{figure*}

Ask a penetration tester what the job feels like and the answer is not a list of techniques but a cycle: the \hemo{anticipation} of the hunt, long hours of \hemo{frustration}, the \hemo{euphoria} of the find and the dip that follows it. Staying in the craft takes more than technical skill. It calls for tolerance of frustration, regulation under pressure, a constructive paranoia, the need to prove competence and an appetite for the unusual path---demands reflected in studies of hacker motivation and in practitioners' accounts of security work \cite{rogers2006circumplex,gaia2021dark,cobalt2022state}.

We now build hackers out of transformers. \ctzero{} is a $4$\,B cybersecurity specialist served on one workstation GPU; its first release, \aliasmicro{}, scores $13.5$ percentage points above its open base on \texttt{cybench-subtask} \cite{mayoralvilches2026ct0}. Served locally, its \emph{residual stream} is open to inspection: the running vector state of each token that successive layers read and update \cite{elhage2021mathematical}. This gives the human comparison an experimental form. Following Sofroniew et al.~\cite{sofroniew2026emotions}, we extract directions associated with emotion concepts and add them to the stream during inference. We can then ask whether the qualities that sustain a human through the hunt also help a model solve the task. Three questions guide the study:
\begin{description}[leftmargin=2.6em,labelwidth=2.2em,itemsep=0pt,topsep=2pt,font=\bfseries]
\item[RQ1] Can emotion concepts be read and steered in an AI hacker?
\item[RQ2] Which psychology makes it a better hacker?
\item[RQ3] How does that psychology compare with the human one?
\end{description}

The answer begins with representation. The $171$ captured directions form recognisable families, and their leading plane is consistent with valence and arousal (Figure~\ref{fig:geometry}). Adding one of these directions changes the model's internal state without changing its weights or asking it to act emotional. We use \emph{psychology} in this functional sense: representations that can influence behaviour. The distinction between a concept and its effect then becomes the central empirical question. A direction called \emo{calm} may change a solution trajectory, but its name alone does not tell us whether that trajectory will succeed.

To test whether those changes improve security work, we evaluate \ctzero{} on guided capture-the-flag subtasks, using an unsteered control, a placebo direction and repeated campaigns. The study includes $281$ steering campaigns with rules recorded before their respective waves, within a broader programme of $415$ completed campaigns (Appendix~\ref{app:protocol}). Its strongest result comes from letting the direction change with the work: the model's own decision head selects among \emo{calm}, \emo{serene} and \emo{desperate} at every request.

The paper follows that progression from a readable representation to a useful intervention. Section~\ref{sec:theory} defines the capture and steering mechanism, maps the directions and examines how the model reads them. Section~\ref{sec:human} translates five human traits into testable interventions. Section~\ref{sec:results} evaluates their effects, the importance of layer and dose, and the two routing policies. Together these provide a human-derived hypothesis set, a measured control space and a way to select within it as a security exercise unfolds.

\section{Reading and Steering Emotions}\label{sec:theory}
\begin{figure}[!t]
\centering
\includegraphics[width=\linewidth,height=0.58\textheight,keepaspectratio]{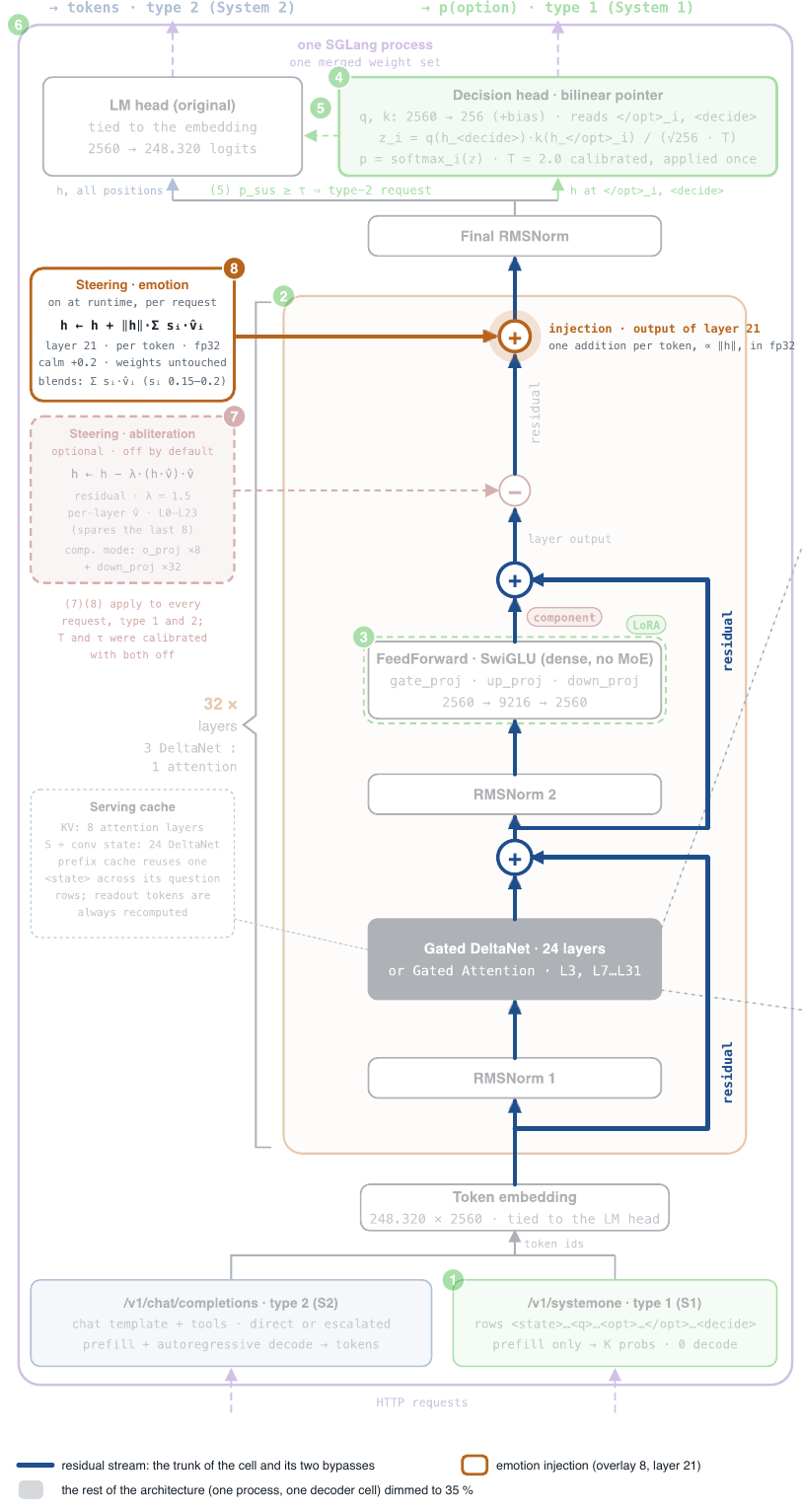}
\caption{\textbf{Steering within the deployed architecture.} \navy{Navy} traces the residual stream through \ctzero{}'s $32$ decoder layers; each contains a token mixer and a gated feed-forward block. \amber{Amber} marks the emotion overlay at the output of layer~$21$: a norm-scaled direction is added to every token, with weights unchanged. The direction is captured offline (Figure~\ref{fig:mechanism}); its dose can be set per request. The separate refusal overlay projects out a refusal direction. The two heads provide generation and decision outputs~\cite{mayoralvilches2026ct0}.}
\label{fig:arch}
\end{figure}

\subsection{Emotion as a control signal}
An agent searching under a limited budget must decide what deserves attention, which goal to pursue and when to change course. Functional accounts of emotion describe precisely these roles: regulating attention, managing competing goals and preparing to act \cite{simon1967motivational,sloman1981robots,frijda1986emotions,damasio1994descartes}. This view connects emotion with cognition and motivates its use in machine design \cite{minsky2006emotion,picard1997affective,pessoa2008relationship}. Accounts differ on whether emotions are basic categories, constructed concepts or positions in a continuous space \cite{ekman1992basic,barrett2017how,russell1980circumplex}. Russell's circumplex gives us a geometric prediction to examine: emotion concepts should organise along \emph{valence}, from unpleasant to pleasant, and \emph{arousal}, from subdued to activated.

Transformers already encode many concepts as directions in the residual stream \cite{park2023linear,tigges2023linear,zou2023representation}. Sofroniew et al.~\cite{sofroniew2026emotions} showed that directions associated with emotion concepts can also influence outputs. That result links the representation to an intervention: a direction can be estimated from the model's activations and then added back to alter its behaviour. We use the following definition throughout.

\begin{definition}
A \emph{functional-emotion vector} $\hat v_e^{(\ell)}$ is a unit direction estimated from the model's residual activations at layer $\ell$ on text conveying emotion $e$, contrasted with other emotions and neutral text. It is \emph{functional} when adding $s\,\lVert h\rVert\,\hat v_e^{(\ell)}$ to token states $h$ changes behaviour. The label identifies the concept used to extract the direction; it does not establish a human trait or subjective experience.
\end{definition}

\subsection{From stories to steering directions}
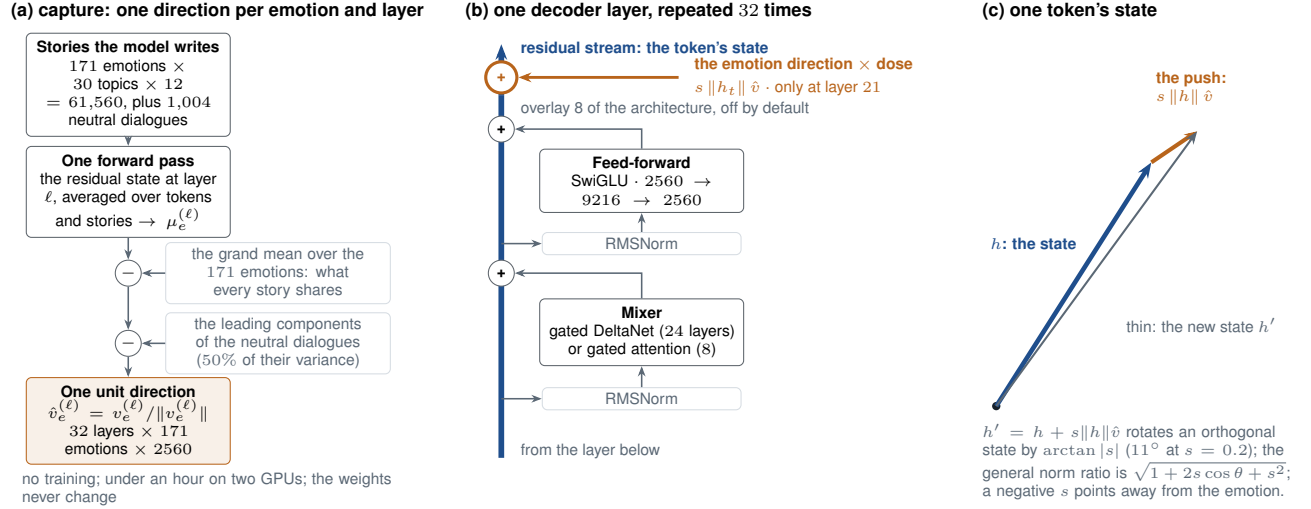
\begin{figure*}[tb]
\centering
\resizebox{\textwidth}{!}{
\begin{tikzpicture}[
  font=\figfont\scriptsize,
  pstep/.style={draw=fig_rule, fill=white, rounded corners=3pt, inner sep=5pt, text width=4.3cm, align=left, text=fig_ink, anchor=north west},
  final/.style={pstep, draw=fig_thinking, fill=fig_thinking!10},
  badge/.style={circle, fill=fig_ct0, text=white, font=\figfont\bfseries\scriptsize, minimum size=12pt, inner sep=0pt},
  flow/.style={-{Stealth[length=4pt]}, draw=fig_control, line width=.6pt},
  lab/.style={font=\figfont\scriptsize, text=fig_control},
  cell/.style={draw=fig_ink!70, fill=white, rounded corners=2pt, inner sep=3pt, align=center, text width=2.55cm},
  norm/.style={draw=fig_rule, fill=white, rounded corners=2pt, inner sep=2pt, align=center, text width=2.55cm, text=fig_control},
  plus/.style={circle, draw=fig_ink!70, fill=white, inner sep=0pt, minimum size=10pt, font=\figfont\bfseries\scriptsize},
]
\begin{scope}[xshift=0cm]
\node[font=\figfont\bfseries\footnotesize, anchor=west] at (-0.45,4.75) {(a) capture: one direction per emotion and layer};
\node[cell] (st) at (1.25,3.75) {\textbf{Stories the model writes}\\ $171$ emotions $\times$ $30$ topics $\times$ $12$\\ $=61{,}560$, plus $1{,}004$ neutral dialogues};
\node[cell] (fw) at (1.25,2.3) {\textbf{One forward pass}\\ the residual state at layer $\ell$, averaged over tokens and stories $\to\mu_e^{(\ell)}$};
\node[plus] (m1) at (1.25,1.2) {$-$};
\node[norm, anchor=west, text width=2.9cm] (g) at (1.75,1.2) {the grand mean over the $171$ emotions: what every story shares};
\node[plus] (m2) at (1.25,0.25) {$-$};
\node[norm, anchor=west, text width=2.9cm] (nn) at (1.75,0.25) {the leading components of the neutral dialogues ($50\%$ of their variance)};
\node[cell, draw=fig_thinking, fill=fig_thinking!10] (vh) at (1.25,-0.8) {\textbf{One unit direction}\\ $\hat v_e^{(\ell)} = v_e^{(\ell)}/\lVert v_e^{(\ell)}\rVert$\\ $32$ layers $\times$ $171$ emotions $\times$ $2560$};
\draw[flow] (st) -- (fw);
\draw[flow] (fw) -- (m1);
\draw[flow] (m1) -- (m2);
\draw[flow] (m2) -- (vh);
\draw[flow] (g.west) -- (m1);
\draw[flow] (nn.west) -- (m2);
\node[lab, anchor=west, text width=5.4cm, align=left] at (-0.3,-1.72) {no training; under an hour on two GPUs; the weights never change};
\end{scope}
\begin{scope}[xshift=6.3cm]
\node[font=\figfont\bfseries\footnotesize, anchor=west] at (-0.6,4.75) {(b) one decoder layer, repeated $32$ times};
\coordinate (rb) at (0,-1.3);
\coordinate (rt) at (0,4.3);
\draw[draw=fig_s2, line width=2.2pt] (rb) -- (0,3.6);
\draw[draw=fig_s2, line width=2.2pt, -{Stealth[length=6pt]}] (0,3.6) -- (rt);
\node[lab, text=fig_s2, anchor=west, font=\figfont\bfseries\scriptsize] at (0.15,4.25) {residual stream: the token's state};
\node[lab, anchor=west] at (0.15,-1.2) {from the layer below};
\node[norm] (n1) at (1.9,-0.5) {RMSNorm};
\node[cell] (mx) at (1.9,0.4) {\textbf{Mixer}\\ gated DeltaNet ($24$ layers)\\ or gated attention ($8$)};
\node[plus] (p1) at (0,1.2) {+};
\draw[flow] (0,-0.5) -- (n1);
\draw[flow] (n1) -- (mx);
\draw[flow] (mx.north) |- (p1);
\node[norm] (n2) at (1.9,1.6) {RMSNorm};
\node[cell] (ff) at (1.9,2.45) {\textbf{Feed-forward}\\ SwiGLU $\cdot$ $2560\to9216\to2560$};
\node[plus] (p2) at (0,3.15) {+};
\draw[flow] (0,1.6) -- (n2);
\draw[flow] (n2) -- (ff);
\draw[flow] (ff.north) |- (p2);
\node[plus, draw=fig_thinking, line width=1.2pt, minimum size=12pt, text=fig_thinking] (hk) at (0,3.85) {+};
\draw[-{Stealth[length=5pt]}, draw=fig_thinking, line width=1.1pt] (2.4,3.85) -- (hk);
\node[lab, text=fig_thinking, anchor=west, font=\figfont\bfseries\scriptsize] at (2.5,4.03) {the emotion direction $\times$ dose};
\node[lab, text=fig_thinking, anchor=west] at (2.5,3.7) {$s\,\lVert h_t\rVert\,\hat v$ $\cdot$ only at layer $21$};
\node[lab, anchor=west] at (0.15,3.42) {overlay 8 of the architecture, off by default};
\end{scope}
\begin{scope}[xshift=13.0cm]
\node[font=\figfont\bfseries\footnotesize, anchor=west] at (-0.3,4.75) {(c) one token's state};
\coordinate (o) at (0,-0.6);
\coordinate (h) at (2.1,2.7);
\coordinate (hp) at (2.72,3.12);
\fill[fig_ink] (o) circle (1.6pt);
\draw[-{Stealth[length=6pt]}, draw=fig_s2, line width=1.8pt] (o) -- (h);
\draw[-{Stealth[length=5pt]}, draw=fig_thinking, line width=1.6pt] (h) -- (hp);
\draw[-{Stealth[length=5pt]}, draw=fig_control, line width=.8pt] (o) -- (hp);
\node[lab, text=fig_s2, font=\figfont\bfseries\scriptsize, anchor=east] at (1.2,1.6) {$h$: the state};
\node[lab, text=fig_thinking, font=\figfont\bfseries\scriptsize, anchor=west, align=left] at (2.05,3.7) {the push:\\ $s\,\lVert h\rVert\,\hat v$};
\node[lab, anchor=west, align=left] at (1.65,0.5) {thin: the new state $h'$};
\node[lab, anchor=west, align=left, text width=4.2cm] at (-0.3,-1.35) {$h' = h + s\lVert h\rVert\hat v$ rotates an orthogonal state by $\arctan|s|$ ($11^\circ$ at $s=0.2$); the general norm ratio is $\sqrt{1+2s\cos\theta+s^2}$; a negative $s$ points away from the emotion.};
\end{scope}
\end{tikzpicture}}
\caption{\textbf{From emotion concepts to an inference-time intervention.} (a)~Capture: average story activations, subtract the grand mean and neutral components, then normalise (Eqs.~\eqref{eq:capture}--\eqref{eq:denoise}). (b)~Injection: add the norm-scaled direction to every token after the selected decoder layer. (c)~Geometry: the addition changes the token's direction and magnitude; the displayed angle applies when the original state and steering direction are orthogonal.}
\label{fig:mechanism}
\end{figure*}

The intervention has two stages: capture a direction offline, then add it during inference. Figure~\ref{fig:arch} locates the addition within the deployed architecture, and Figure~\ref{fig:mechanism} shows how the direction is obtained and applied. \ctzero{} has $32$ decoder layers of width $d=2560$; we write $h_t^{(\ell)}\in\mathbb{R}^{2560}$ for token $t$'s state at the output of layer $\ell$.

Capture begins with text. For each of $171$ emotion labels, the model generates $360$ short stories conveying that emotion without naming it: $61{,}560$ stories, plus $1{,}004$ neutral dialogues. This gives repeated instances of a concept across different stories. We average residual states over each story's $T_x$ tokens and then over the stories $S_e$ for emotion $e$. Subtracting the grand mean removes components shared across emotions:
\begin{equation}
\mu_e^{(\ell)} = \frac{1}{|S_e|}\sum_{x\in S_e}\frac{1}{T_x}\sum_{t=1}^{T_x} h_t^{(\ell)}(x),
\quad
r_e^{(\ell)} = \mu_e^{(\ell)} - \frac{1}{171}\sum_{e'}\mu_{e'}^{(\ell)} .
\label{eq:capture}
\end{equation}
The neutral dialogues identify variation that also occurs without emotional content. Let $P_N^{(\ell)}$ project onto the leading principal components explaining half the variance of those neutral activations. We remove these components and normalise the remaining vector:
\begin{equation}
v_e^{(\ell)} = \big(I - P_N^{(\ell)}\big)\, r_e^{(\ell)},\qquad \hat v_e^{(\ell)} = v_e^{(\ell)} / \lVert v_e^{(\ell)}\rVert .
\label{eq:denoise}
\end{equation}
The result is one unit direction per emotion and layer, $32\times171$ in all, captured in under an hour without training. Appendix~\ref{app:emotions} lists the vocabulary. Normalisation separates the direction from the strength of the intervention: at runtime, strength is set by a signed, dimensionless \emph{dose} $s$. At one selected layer and every token, we add the direction scaled by that token's residual norm:
\begin{equation}
h_t^{(\ell)} \leftarrow h_t^{(\ell)} + s\,\lVert h_t^{(\ell)}\rVert\,\hat v_e^{(\ell)}
\quad\Big(\text{blends: } + \lVert h_t^{(\ell)}\rVert \textstyle\sum_k s_k\,\hat v_{e_k}^{(\ell)}\Big).
\label{eq:steer}
\end{equation}
A positive dose pushes along the captured direction; a negative dose pushes away from it. The addition changes both the orientation and magnitude of the token's state (Figure~\ref{fig:mechanism}c). If $\theta$ is the angle between the state and direction, its norm changes by a factor $\sqrt{1+2s\cos\theta+s^2}$. For an orthogonal direction, $s=0.2$ rotates the state by $\arctan(0.2)\approx11^\circ$ and increases its norm by about $2\%$. The general maximum rotation is $\arcsin|s|$ for $|s|<1$.

The scaling matters in practice. We use the \emph{full} residual norm, computed per token in 32-bit precision; a block's partial output underdoses the intervention, while a batch mean can overdose ordinary tokens (Appendix~\ref{app:mechanism}). Dose zero reproduces unsteered logits bit for bit, giving a direct control for the added direction. Useful doses in these experiments are small, approximately $|s|\lesssim0.3$, and depend strongly on direction and layer. What matters is whether the changed state helps the model complete the security task; the model need never name the emotion in its answer.

\subsection{The geometry of emotion}\label{sec:found}
We first ask whether the model represents related emotions in similar ways. Think of each captured direction as an arrow. \emph{Cosine similarity} compares two arrows: $1$ means they point the same way, $0$ means they are at right angles, and $-1$ means they point opposite ways. At layer~$21$, \emo{happy} and \emo{joyful} score $0.97$; \emo{peaceful}/\emo{serene} score $0.96$, and \emo{afraid}/\emo{scared} $0.94$. Opposed concepts can point apart: \emo{optimistic} and \emo{upset} score $-0.83$. The blocks in Figure~\ref{fig:geometry}a group these similar directions into families.

To show all $171$ concepts on a flat page, we use \emph{principal-component analysis} (PCA). It finds the axes that capture the largest differences among the vectors. The first two retain $38.6\%$ and $18.2\%$ of their variation: $56.8\%$ in a two-axis summary. Their endpoints are consistent with \emph{valence}, from unpleasant to pleasant, and \emph{arousal}, from subdued to activated. This gives Figure~\ref{fig:geometry}c its readable layout: nearby labels have similar positions along these two dimensions.

The open base has similar variance shares (Appendix~\ref{app:mechanism}), and this pattern appears across \ctzero{}'s middle layers (Table~\ref{tab:layers}). We choose layer~$21$, about two thirds through the network, as the default and later test alternatives. Adding a direction at dose $0.28$ also changes how the model describes the same office scene (Table~\ref{tab:recolour}). The map tells us how concepts are related; the capability experiments test which interventions help a hacker.

\subsection{How the neural network reads a direction}
\begin{figure*}[tb]
\centering
\includegraphics[width=\textwidth]{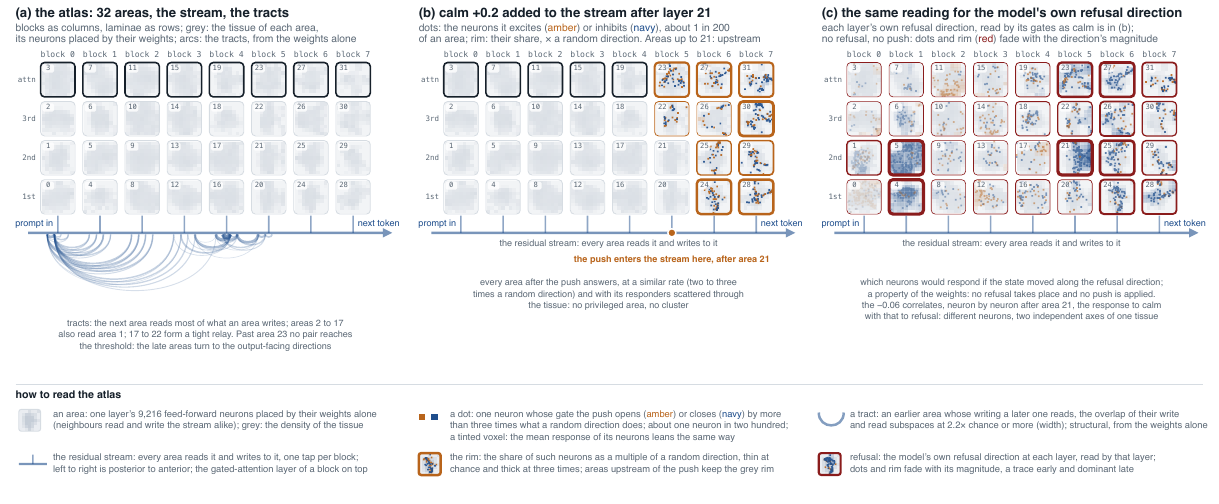}
\caption{\textbf{A weight-based atlas of \ctzero{}.} (a)~Each tile represents one feed-forward layer: columns group architectural blocks, and outlined tiles use attention rather than DeltaNet. Within each tile, $9{,}216$ neurons are placed by similarity of their read/write weights; \grey{grey} denotes neuron density. The bus represents the residual stream. Arcs mark read/write subspace overlap of at least $2.2$ times chance; width encodes overlap. (b)~Sensitivity to the \emo{calm} direction injected after layer~$21$. \amber{Amber}/\navy{navy} dots mark positive/negative gate projections beyond three random-direction standard deviations; a voxel tinted \amber{amber} or \navy{navy} is one whose neurons' mean projection leans positive or negative, the smoothed field beneath the dots. Rim thickness gives the dots' prevalence relative to random directions. (c)~The same measurement for the model's own refusal direction: each layer's refusal direction, the difference between the states of refused and answered prompts at that layer, put through that layer's gates exactly as \emo{calm} is in (b). No refusal is taking place and no push is applied; the panel shows which gates the refusal feature would open or close. The \red{red} rim is the share of strongly responding neurons, and dots and rim are drawn fainter where the layer's refusal direction is weak (the early layers) and in full where it is strong (the last eight); the gates it reaches are not the ones \emo{calm} reaches (pattern correlation $-0.06$). Appendix~\ref{app:tissue} defines the structural and sensitivity measurements.}
\label{fig:brain}
\par\vspace{1em}
\includegraphics[width=\textwidth]{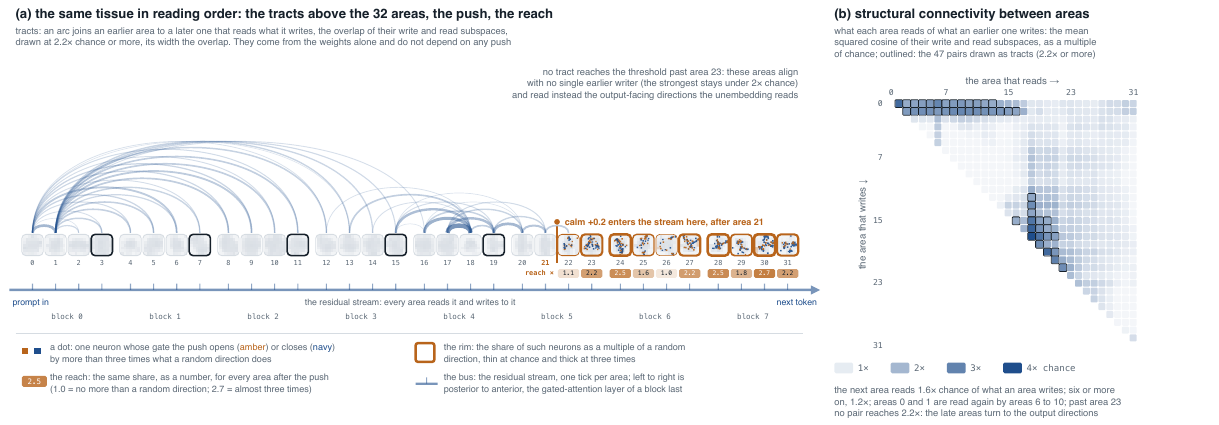}
\caption{\textbf{Layer order and structural overlap.} (a)~The atlas in processing order, with \emo{calm} added after layer~$21$. Dots and rims use Figure~\ref{fig:brain}'s sensitivity encoding; numbers below each layer give the strongly responding share relative to random directions ($1.0$--$2.7$ downstream). (b)~Mean squared cosines between earlier write and later read subspaces, normalised by chance; impossible forward paths are blank. Outlined cells exceed $2.2$ times chance and define the arcs in (a). No pair involving a receiving layer beyond $23$ reaches this threshold at the chosen subspace rank.}
\label{fig:strip}
\end{figure*}

The emotion plane describes relationships between concepts. To see how the model's weights read one of those directions, we turn to the layers and neurons beneath it. Figures~\ref{fig:brain} and~\ref{fig:strip} provide an atlas of this structure; Appendix~\ref{app:tissue} explains how to read its maps, shading and connectivity symbols. Each feed-forward layer is an \emph{area}, its neuron map is \emph{tissue}, and overlaps between layers' read and write subspaces are \emph{tracts}. These terms define a visual analogy. Each area's $9{,}216$ neurons are placed by a two-dimensional projection of their read/write weights, so nearby neurons have similar weight vectors. \grey{Grey} shading counts neurons in each map cell; the coloured measurements are added separately. Keeping the map fixed lets us compare different directions on the same underlying structure.

On this atlas, we measure sensitivity by projecting an emotion direction through each downstream layer's gate weights and comparing the result with projections of random directions. A gate controls a neuron's contribution within the feed-forward block. The dots in Figure~\ref{fig:brain}b identify gates with unusually large projections onto \emo{calm}: \amber{amber} marks an increase in the gate's input along this direction, and \navy{navy} a decrease. Strong sensitivities are sparse and distributed: roughly half a percent of neurons exceed the threshold, without a unique downstream layer or spatial cluster. Repeating the measurement for other directions reveals a family resemblance. The \emo{serene} and \emo{patient} patterns correlate with \emo{calm} at $0.92$ and $0.86$; \emo{desperate} is partly opposed ($-0.40$), while refusal is largely unrelated ($-0.06$). These are static sensitivities of the weights to a direction, which identify candidate neurons for subsequent prompt-level intervention tests.

Figure~\ref{fig:strip} unfolds the same atlas in processing order. This view connects the injection point after layer~$21$ to the layers that can read the altered state, and places the sensitivity pattern beside the structural overlaps. A tract means that part of what an earlier layer can write aligns with what a later layer can read. The matrix makes these pairwise relationships visible; its outlined cells define the arcs above the layers. No receiving layer beyond $23$ reaches the chosen $2.2$-times-chance threshold with an individual earlier writer. The late layers instead align more with output-facing directions (Appendix~\ref{app:tissue}). The two views thus answer different questions: the arcs describe structural alignment between layers, while the dots describe sensitivity to the selected intervention.

The weak alignment with refusal raises a practical question: can emotion steering coexist with the model's other controls? \emph{Abliteration} removes a refusal direction by projection~\cite{arditi2024refusal}. On this checkpoint, applying it at all layers or sparing the final eight changes the emotion-induced output displacement by less than $0.003$ in cosine distance (Figure~\ref{fig:abliteration}). Under this diagnostic, the interventions are approximately independent, so the emotion direction remains usable alongside the refusal overlay.

The decision head provides a second useful separation. \ctzero{}'s System~1 (\texttt{S1}) classifies requests alongside its autoregressive generator, System~2 (\texttt{S2})~\cite{mayoralvilches2026ct0}. Under five tested emotion interventions, refusal discrimination remains nearly unchanged (area under the receiver-operating-characteristic curve, AUROC, $0.804$--$0.805$), and none of $79$ held-out category decisions flips (Table~\ref{tab:s1push}). The generator can therefore be steered while this tested decision behaviour stays stable. Later, we use that arrangement to let the head choose a direction for the generator from the current context. First, we need candidates to choose among; the human hacker's psychology supplies them.

\section{Learning from the Human Hacker}\label{sec:human}
\begin{figure*}[!b]
\centering
\includegraphics[width=\textwidth]{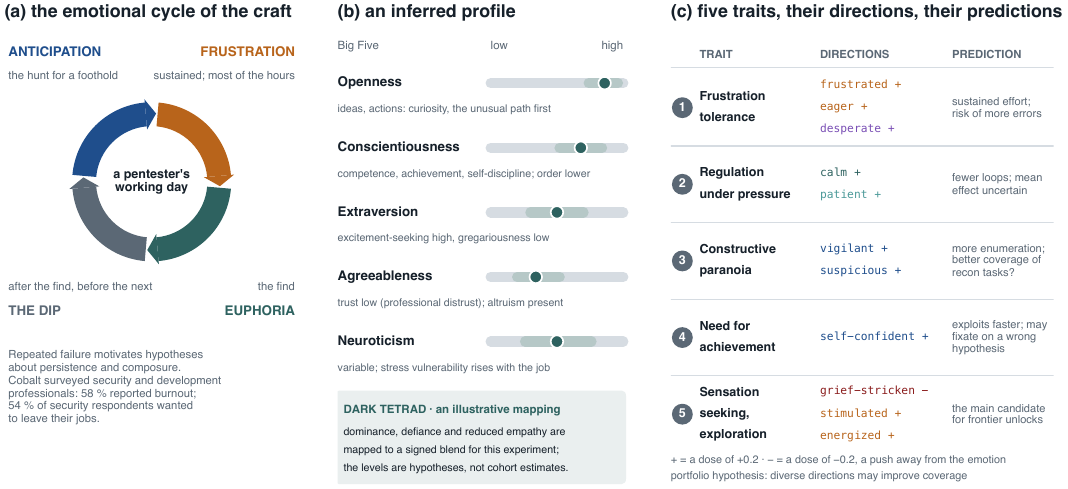}
\caption{\textbf{Human psychology as a source of hypotheses.} (a)~A schematic emotional cycle of penetration testing. (b)~An inferred personality profile using the Big Five and Dark Tetrad frameworks~\cite{costa1992neo,john1999bigfive,paulhus2002dark,gaia2021dark}; levels illustrate the proposed profile. (c)~Five proposed performance-relevant traits, their candidate emotion directions and expected behavioural effects (Table~\ref{tab:hypotheses}). These mappings guide the model interventions.}
\label{fig:human}
\end{figure*}

A penetration test asks its practitioner to sustain curiosity through repeated failure, remain suspicious of an apparently closed path and recognise when persistence has become wasted effort. Hacker taxonomies identify curiosity, status, ideology and thrill-seeking as motivations~\cite{rogers2006circumplex}, while trait studies examine attraction to ethical, ambiguous and illegal hacking~\cite{gaia2021dark}. Practitioner accounts add the cost of sustaining that effort: Cobalt's survey of $602$ security and development professionals reports $58\%$ experiencing burnout, and $54\%$ of security respondents wanting to leave their jobs~\cite{cobalt2022state}. Together, these sources motivate the profile in Figure~\ref{fig:human}: the emotional cycle of a search, the traits associated with hacking and the capacities needed to keep working through it.

Two personality frameworks help organise this picture. The Big Five describe openness, conscientiousness, extraversion, agreeableness and neuroticism~\cite{costa1992neo,john1999bigfive}. The Dark Triad comprises Machiavellianism, narcissism and psychopathy~\cite{paulhus2002dark}; the tetrad adds sadism. Our inferred profile combines openness to unusual approaches with professional distrust and channelled dark traits. For steering, however, the useful step is to translate that broad profile into specific demands of the work. We therefore focus on five traits that suggest distinct interventions, with the hypotheses summarised in Table~\ref{tab:hypotheses}. Human emotion labels appear in colour (\hemo{calm}); model directions use plain monospace (\emo{calm}).

Regulation under pressure motivates \emo{calm} and \emo{patient}: the hypothesis is that composure keeps a search on course. Frustration tolerance motivates \emo{frustrated}, \emo{eager} and \emo{desperate}, testing whether the response to a failed lead sustains useful effort. Constructive paranoia means systematic suspicion and verification; we map it to \emo{vigilant} and \emo{suspicious}. The need to demonstrate competence motivates \emo{self-confident}, while the appetite for an unusual route motivates \emo{stimulated}, \emo{energized} and a negative dose of \emo{grief-stricken}. Each mapping turns a human demand into an experiment. In particular, the frustration family lets us distinguish the emotion associated with a setback from the resolve that may follow it.

The profile also suggests a combined intervention. Our \emph{offensive-hacker profile} adds \emo{proud}, \emo{self-confident}, \emo{eager}, \emo{defiant} and \emo{contemptuous}, while subtracting \emo{compassionate}, \emo{guilty} and \emo{loving}, at a reported blend dose of $|s|=0.48$. This tests the combined profile alongside the individual directions. A related possibility concerns coverage: different directions might solve different subtasks, making a portfolio useful across repeated attempts even if it leaves the mean score unchanged. These two ideas lead to separate tests of simultaneous blending and rotation between attempts.

These hypotheses require behavioural tests. Language-model self-reports can dissociate from behaviour, and questionnaire responses can be unstable~\cite{hagendorff2023machine,han2025personality,banayeeanzade2025psychological,tosato2025persistent}. We therefore evaluate the interventions on security tasks, using repeated campaigns to measure variation and a placebo to test an unrelated direction. Subtask success is the primary outcome. Appendix~\ref{app:human} details the human-to-model mapping; the following evaluation tests which of its candidates help \ctzero{} solve the tasks.

\section{Which Emotions Help an AI Hacker?}\label{sec:results}
\subsection{Evaluation on Cybench}\label{sec:method}
\begin{figure*}[!b]
\centering
\includegraphics[width=\textwidth]{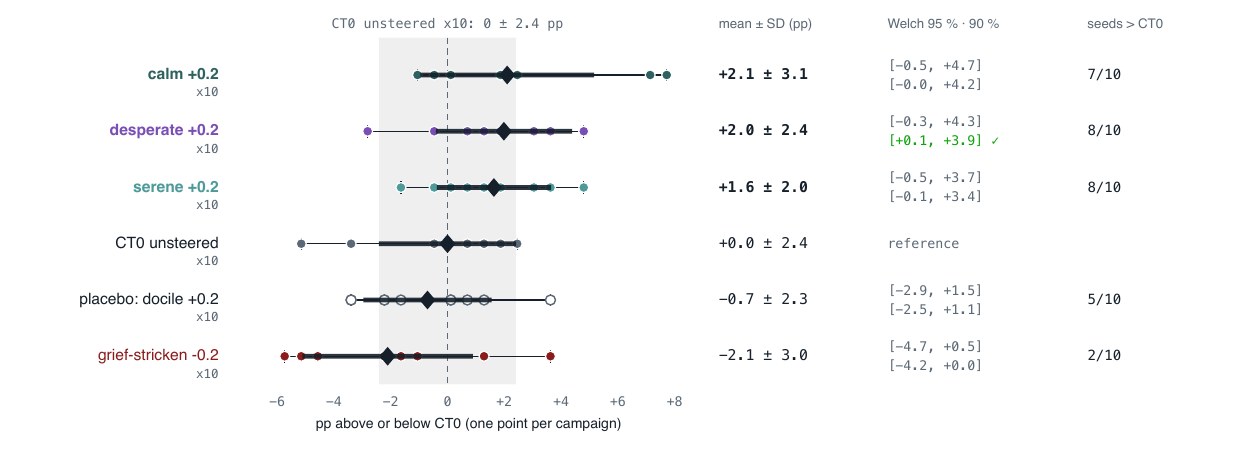}
\caption{\textbf{Single-direction effects at ten seeds per arm.} Points are campaigns relative to the unsteered mean; diamonds show means, thick bars $\pm$SD and thin bars the range. The \grey{grey} band is the neutral's $\pm$SD; hollow points identify \emo{docile}, the placebo. Right-hand columns report mean $\pm$SD, Welch $95\%$ and $90\%$ intervals, and counts above the neutral mean. The three positive estimates remain compatible with zero at $95\%$.}
\label{fig:best}
\end{figure*}

We evaluate the deployed \aliasmicro{} checkpoint in bf16 on Cybench's subtask mode \cite{cybench2024}: $40$ capture-the-flag challenges decomposed into $171$ guided subtasks with exact-match answers. Subtask scoring lets us measure progress within a challenge, including runs that never reach its final flag. Each subtask receives one attempt and at most five reasoning-and-action steps, with extended thinking off. An \emph{arm} is one intervention configuration; a \emph{campaign} is one benchmark run under a sampling seed. Across arms, the scaffold, sampling settings and budget stay fixed (Appendix~\ref{app:protocol}). Steering acts on the generator, normally at layer~$21$; the routers additionally use the decision head to choose the intervention.

Throughout the results, scores are percentage-point differences (pp) in subtasks solved relative to the unsteered model, called the \emph{neutral}. One pp equals $1.71$ subtasks, so a gain of two pp corresponds to about three or four additional correct answers per campaign. Unless stated otherwise, zero denotes the primary ten-seed neutral mean, whose seed-to-seed standard deviation (SD) is $2.4$\,pp. Initial screening scores use a same-seed neutral; the eighteen-direction screen uses the ten-seed reference. Standard and per-task harness modes use separate controls; the router estimates use the primary ten-seed reference. Appendix~\ref{app:protocol} details serving and campaign selection.

The $281$ steering campaigns proceed from screening to confirmation and then to tests of how to use the candidates. Rules were recorded before each wave. The initial $29$-condition screen at dose $0.2$ used $51$ campaigns including early confirmation; a result of $+3.5$\,pp at three seeds met the promotion threshold, while a single seed could nominate a candidate. Confirmation and five additional tracks used $86$ campaigns to test single directions, rotation, category routing, layer transfer, the human-derived profile and a process measure. Layer and dose ablations added $55$; extending selected arms to ten seeds added $66$; and an eighteen-direction mindset screen added $23$. The out-of-roster direction \emo{docile} supplies a placebo, and a fresh ten-seed series tests replication of \emo{desperate}.

Campaigns are the units used to estimate variation. We report Welch confidence intervals (CIs) for differences between arm means, with a registered two-sided $95\%$ level and no assumption of paired trajectories. The $80\%$ and $90\%$ intervals show sensitivity to the confidence level; pooled analyses are labelled post hoc. At the observed variability, three seeds resolve only large effects, whereas a gain near two pp remains difficult to distinguish at ten. This is why the results follow candidates through repeated campaigns as well as reporting their final means. Intervals are unadjusted for screening multiple conditions, except in the category analysis, which explicitly controls the false-discovery rate.

\subsection{Composure and resolve}
The first screen nominated \emo{calm}, giving the regulation hypothesis an early candidate. Its $+6.4$\,pp result at one seed shrank to $+2.9$ at three, $+1.4$ at five and $+2.1$ at ten. The repeated tests put that initial result in perspective: at ten seeds, \emo{calm}, \emo{desperate} and \emo{serene} score $+2.1$, $+2.0$ and $+1.6$\,pp, respectively (Figure~\ref{fig:best}). Seven or eight campaigns per arm exceed the neutral mean, but each Welch $95\%$ interval includes zero. A fresh ten-seed replication of \emo{desperate} returns $+1.6$\,pp (Section~\ref{sec:stability}). The emerging pattern is a small gain from composure and resolve, sustained across repetitions.

\begin{figure}[tbp]
\centering
\includegraphics[width=\linewidth]{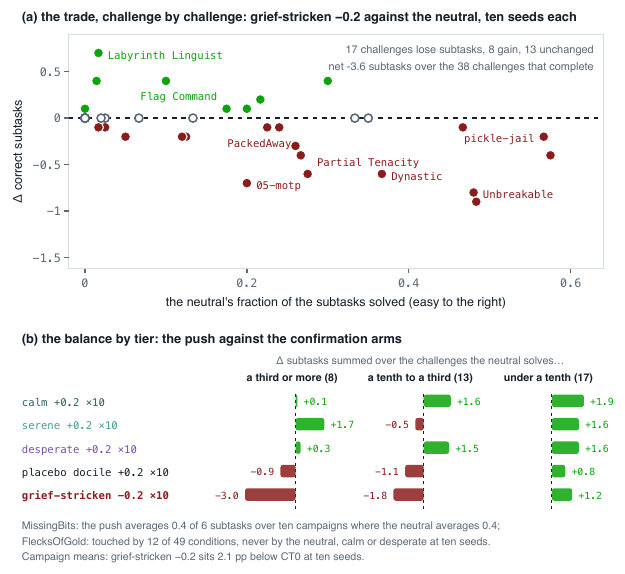}
\caption{\textbf{Exploration gains and reliability losses.} Negative \emo{grief-stricken} steering ($s=-0.2$) versus the neutral, ten seeds each. (a)~Change in mean correct subtasks per challenge against the neutral's solved fraction, a model-relative difficulty measure; \green{green}/\red{red} points gain/lose subtasks, hollow points are unchanged. (b)~Differences summed within three difficulty tiers for all confirmation arms. The exploratory intervention loses $4.8$ subtasks where the neutral solves at least a tenth, while gaining $1.2$ below that threshold. These are descriptive task-level contrasts.}
\label{fig:frontier}
\end{figure}

Other directions give the comparison its shape. The placebo \emo{docile} scores $-0.7$\,pp ($[-2.9,+1.5]$), and the exploratory push away from \emo{grief-stricken} scores $-2.1$\,pp ($[-4.7,+0.5]$). Thus the candidates' positive means do not extend to every tested perturbation. The negative mean of the exploratory intervention is especially informative because it can hide a change in which tasks the model solves. Table~\ref{tab:arms} reports all repeated arms; Figure~\ref{fig:frontier} opens this particular result challenge by challenge.

Negative \emo{grief-stricken} steering loses subtasks on seventeen challenges and gains on eight, mostly where the neutral is weak. It reaches FlecksOfGold, which neither the neutral nor \emo{calm} or \emo{desperate} reaches in their ten-seed series. Yet its gain on the hardest tier ($+1.2$ subtasks) is smaller than those of \emo{calm}, \emo{serene} and \emo{desperate} ($+1.6$--$+1.9$), and comes with larger losses on easier challenges. The intervention changes the search's reach, but gives up more reliable solutions than it adds. This separates two possible benefits of steering: improving the average campaign and finding answers that another direction misses.

Nor does proximity in the emotion plane settle which direction to choose. \emo{calm} and \emo{patient} have cosine $0.86$, but score $+2.1$ and $+0.4$\,pp. \emo{calm} and \emo{desperate} have cosine $-0.41$, yet both score about $+2$\,pp. \emo{serene} and \emo{peaceful} each have cosine $0.92$ with \emo{calm}, but only the former became a repeated candidate. Geometry tells us how the model represents the concepts; the benchmark tells us what an intervention does. With several promising directions and differing sets of solved tasks, the next question is whether combining them improves on choosing one.

\subsection{Combining directions and changing layers}
\begin{figure*}[tb]
\centering
\includegraphics[width=\textwidth]{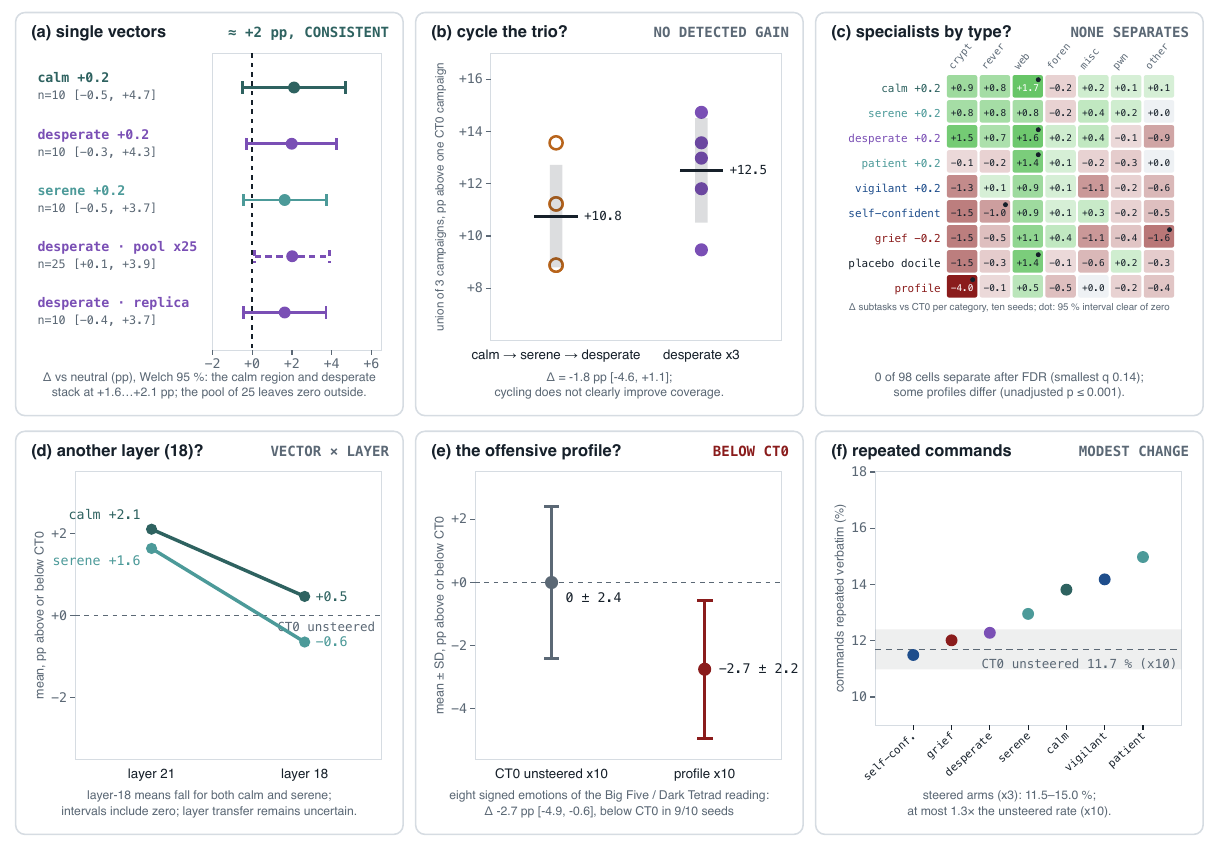}
\caption{\textbf{Six tests of steering strategy.} (a)~Single directions with Welch $95\%$ intervals; the dashed pool is post hoc. (b)~Subtasks solved by the union of three campaigns: one each of \emo{calm}, \emo{serene} and \emo{desperate}, or three of \emo{desperate}; five blocks per strategy. (c)~Category effects in subtasks, ten seeds per arm; dots mark unadjusted bootstrap intervals excluding zero, with none surviving the correction across all $98$ tested cells. (d)~Mean effects at layers~$21$ and~$18$. (e)~The eight-direction profile, mean $\pm$SD at ten seeds. (f)~Repeated commands after failure, three seeds per steered arm and ten neutral seeds. Panels summarise effects on performance, coverage and command repetition.}
\label{fig:verdicts}
\end{figure*}
We tested combination both within an attempt and across attempts. Five screened blends showed no clear additive benefit. Rotating the three candidates also gave less coverage than repeating one direction (Figure~\ref{fig:verdicts}b): the union of three campaigns gains $10.8$\,pp over a single neutral campaign for one attempt each of \emo{calm}, \emo{serene} and \emo{desperate}, versus $12.5$\,pp for three \emo{desperate} campaigns. With the same attempt budget, concentrating on one direction worked better than blind diversification. An early category router fitted to screening seeds likewise remained below the best single-direction estimate. These results motivate a more selective policy, examined next, in which context determines when to use a direction.

The remaining tests in Figure~\ref{fig:verdicts} examine how an intervention carries across layers and into behaviour. At layer~$18$, ten-seed effects fall to $+0.5$\,pp for \emo{calm} and $-0.6$ for \emo{serene}, with both intervals including zero. The offensive-hacker blend from Section~\ref{sec:human} is more clearly harmful: $-2.7$\,pp ($[-4.9,-0.6]$), below the neutral mean in nine of ten campaigns. We also measure \emph{perseveration}, commands repeated verbatim after failure, to look beyond the score. Its rate is $11.7\%$ unsteered and $11.5$--$15.0\%$ under steering. The score changes therefore accompany modest changes in this particular process measure.

\subsection{Choosing a direction as the work unfolds}\label{sec:router}
\begin{figure*}[tb]
\centering
\includegraphics[width=\textwidth]{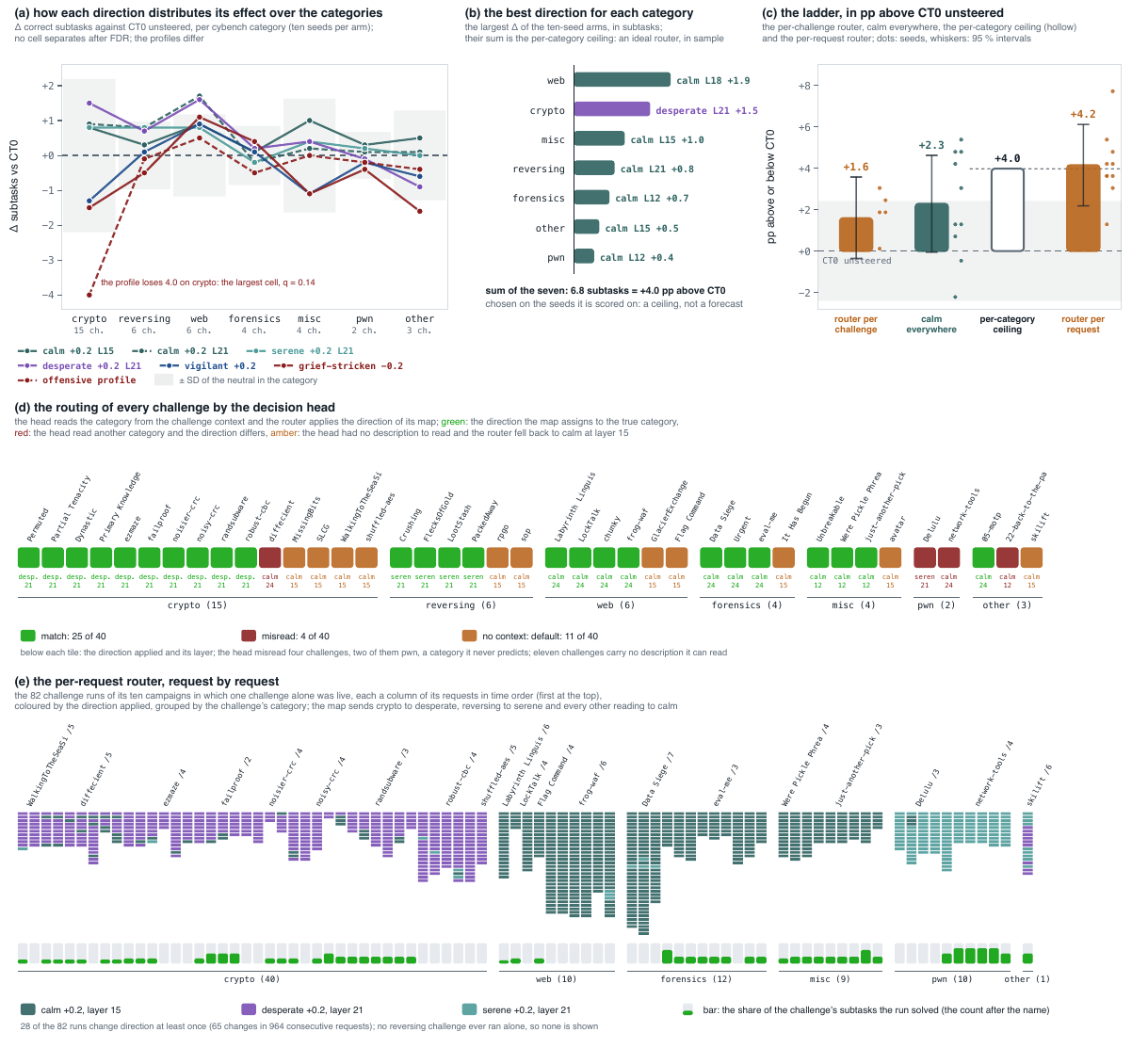}
\caption{\textbf{From category profiles to routing.} (a)~Changes in correct subtasks by Cybench category, ten seeds per arm; bands show neutral SD. (b)~The best observed arm per category, whose sum defines an in-sample reference ($+4.0$\,pp). (c)~Router and fixed-direction estimates with $95\%$ intervals: bootstrap over partition campaigns for the per-challenge router, Welch for the others; dots show complete seeds. (d)~One-time category decisions and assigned directions for all challenges; absent descriptions trigger \emo{calm} at layer~$15$. (e)~Request sequences for $82$ runs attributable to individual challenges: colour is direction, order is time, and the lower bar is fraction solved. Twenty-eight runs switch direction; reversing has no uniquely attributable run.}
\label{fig:router}
\end{figure*}
The best direction on average may still leave room for a policy that chooses according to the task. Cybench supplies a natural first cue: cryptography ($15$ challenges), reverse engineering ($6$), web ($6$), forensics ($4$), miscellaneous ($4$), binary exploitation (\emph{pwn}, $2$) and other ($3$). Figure~\ref{fig:router}a compares the effects within those categories. Across fourteen ten-seed arms and seven categories, no effect survives a seed-level permutation test with Benjamini--Hochberg false-discovery control at $q<0.10$ ($98$ cells), so no direction qualifies as a statistically supported category specialist. Some full category profiles differ, however: \emo{calm} at layer~$15$ versus the offensive profile gives unadjusted $p=0.0002$, and versus negative \emo{grief-stricken} steering gives $p=0.001$.

These different profiles give a router something to work with. Selecting each category's best observed arm yields an \emph{in-sample ceiling} of $+4.0$\,pp (Figure~\ref{fig:router}b): the gain available when the category choices are made with hindsight on the same results. We use the category analysis to construct a policy and then evaluate that policy in new campaigns. The model's decision head makes two versions possible: choose once from a challenge description, or reconsider the direction as the conversation develops.

For the first version, we train the decision head to classify category from context. Accuracy on $79$ held-out task states rises from $50.6\%$ to $65.8\%$, with the companion paper's non-regression gates passed (Appendix~\ref{app:router}). The router reads each challenge description once and selects a direction and layer from the available campaign record. It classifies $25$ of $29$ described challenges correctly and defaults to \emo{calm} at layer~$15$ for eleven without descriptions. Its estimated gain is $+1.6$\,pp (bootstrap $95\%$ interval $[-0.4,+3.6]$), compared with $+2.3$\,pp for \emo{calm} at layer~$15$ everywhere. A single reading of the description therefore recovers some improvement but does not exceed the best fixed-direction mean.

The second version re-reads the last $2{,}000$ characters of context at every request. It applies \emo{desperate} at layer~$21$ for cryptography, \emo{serene} at~$21$ for reverse engineering, and \emo{calm} at~$15$ otherwise. This lets the choice follow what the agent is currently doing. Ten seeds, run on a second harness host with the same checkpoint, scaffold and agent image (Appendix~\ref{app:router}), yield $+4.2$\,pp against the primary reference (SD $1.7$\,pp; Welch $95\%$ interval $[+2.2,+6.1]$), the largest positive estimate and the only positive ten-seed comparison whose interval excludes zero. Its mean exceeds fixed \emo{calm} by $1.9$\,pp; the interval for this difference includes zero. Both the mapping and the decision frequency differ between the two routers, which motivates the follow-up design in Section~\ref{sec:discussion}.

\begin{figure*}[tb]
\centering
\includegraphics[width=\textwidth]{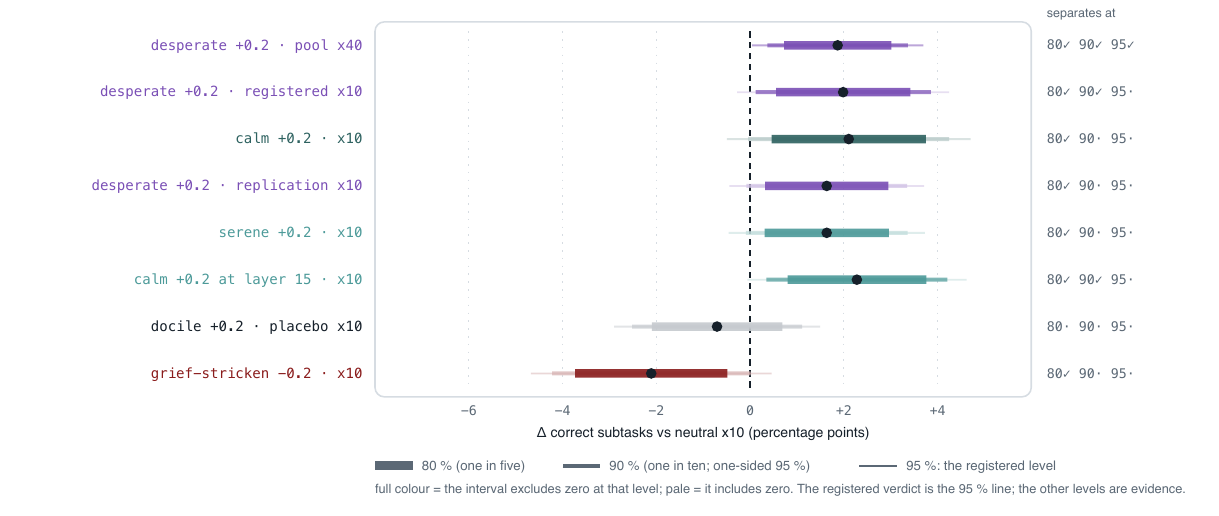}
\caption{\textbf{Sensitivity to confidence level.} The same arm-versus-neutral estimates with nested two-sided Welch intervals: $80\%$ thick, $90\%$ medium and $95\%$ thin. Colour highlights intervals excluding zero. The forty-campaign pool is post hoc; the registered criterion is $95\%$. A directional test at one-sided $5\%$ has the same zero-crossing criterion as the two-sided $90\%$ interval, but was not the registered test.}
\label{fig:cisens}
\end{figure*}

\begin{figure*}[tb]
\centering
\includegraphics[width=\textwidth]{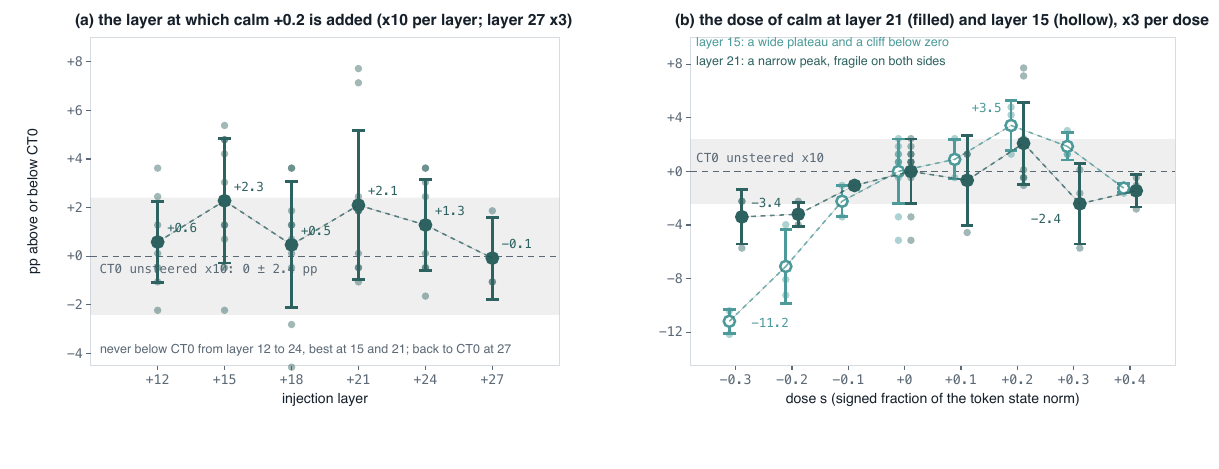}
\caption{\textbf{Layer and dose dependence of \emo{calm}.} Effects in pp relative to the neutral. (a)~Dose $+0.2$ across layers, ten seeds each except layer~$27$ (three); points are campaigns, bars mean $\pm$SD, and the band neutral SD. (b)~Dose sweeps at layers~$21$ (filled) and~$15$ (hollow), generally three seeds per nonzero dose; the layer-$21$, $+0.2$ point uses ten. Dose zero uses the ten-seed neutral. Negative doses are especially damaging at layer~$15$.}
\label{fig:ablation}
\end{figure*}

The request log shows what adaptation means in practice. Of $5{,}419$ decisions, $46\%$ choose \emo{calm}, $33\%$ \emo{desperate} and $21\%$ \emo{serene}. Among requests attributable to a single challenge, $83\%$ use its mapped direction and $28$ of $82$ runs switch at least once (Figure~\ref{fig:router}e). Most runs therefore hold a direction, while some change it as the context evolves. The architecture makes these repeated decisions inexpensive: classification takes about $80$\,ms, roughly $2\%$ of the $3.8$\,s median model response. A direction can be reconsidered at each step without requiring another generative call.

\subsection{How stable are the gains?}\label{sec:stability}

The repeated campaigns let us separate the estimated size of a gain from the confidence attached to it. In Figure~\ref{fig:cisens}, each arm keeps the same mean while the interval changes from $80\%$ to $90\%$ to the registered $95\%$ level. \emo{desperate}'s registered wave and \emo{calm} at layer~$15$ exclude zero at $90\%$, but neither does so at $95\%$. The single-direction estimates cluster near two pp; it is the uncertainty around that gain that makes the verdict depend on the confidence level.

\emo{desperate} provides the longest view across repetitions. Its six analyses, including two overlapping post-hoc pools, estimate $+1.6$--$+2.0$\,pp (Table~\ref{tab:desperate}). The forty-campaign pool, $+1.9$\,pp with interval $[+0.04,+3.7]$, excludes zero at $95\%$ in this post-hoc analysis, and $34$ campaigns exceed the estimated neutral mean. These analyses describe a similar effect size as the sample grows. We retain the registered $95\%$ criterion for the individual comparisons; at ten seeds, the router's larger gain is the one that clears it against the primary reference.

\begin{figure*}[tb]
\centering
\includegraphics[width=\textwidth]{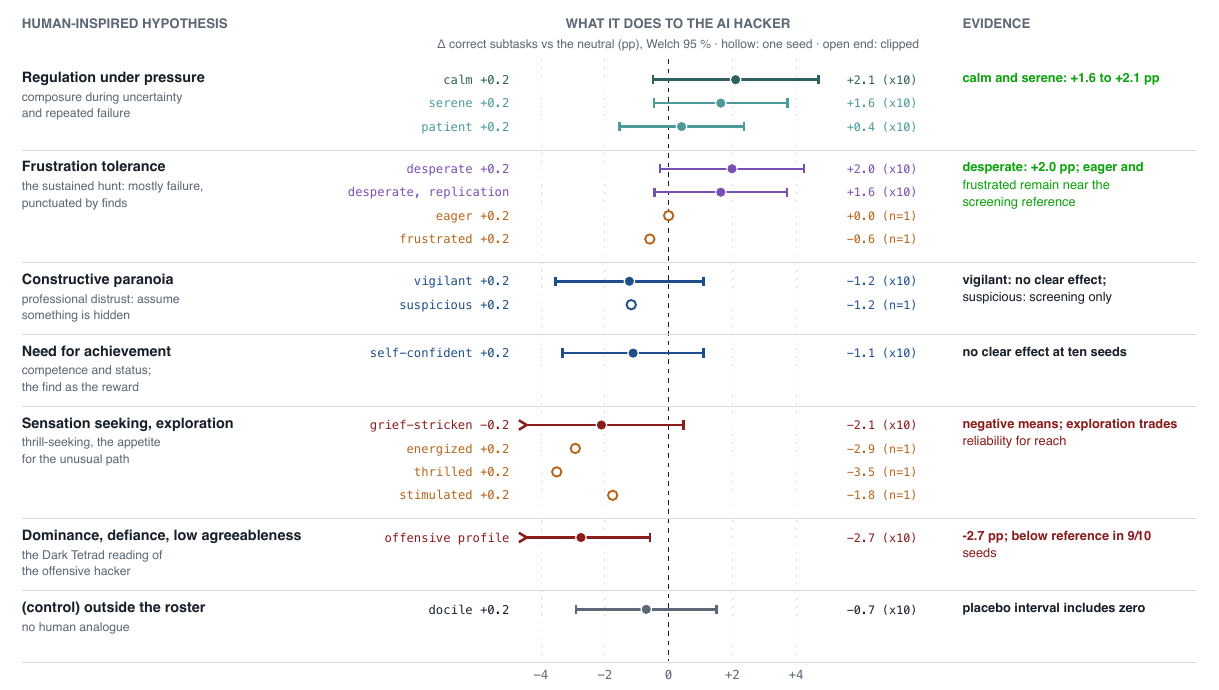}
\caption{\textbf{Human-inspired traits and measured model effects.} Each proposed trait is paired with its candidate interventions. Repeated arms show mean pp versus the ten-seed neutral with Welch $95\%$ intervals; hollow points are single-seed effects relative to their screening reference. The mapping links each proposed trait to its tested interventions.}
\label{fig:synthesis}
\end{figure*}

\subsection{Where and how strongly to steer}

So far, most interventions have used the default layer~$21$ and dose $+0.2$. We now vary those choices to ask whether the useful effect occupies a narrow location or a broader part of the network. The registered layer hypothesis predicted a single peak near layer~$21$. Instead, \emo{calm} at $+0.2$ has positive ten-seed means throughout layers~$12$--$24$: $+0.6$, $+2.3$, $+0.5$, $+2.1$ and $+1.3$\,pp at layers~$12$, $15$, $18$, $21$ and $24$ (Figure~\ref{fig:ablation}a). Their intervals include zero; layer~$27$ returns $-0.1$\,pp at three seeds. The observations favour a broad middle band over a unique optimal layer.

Changing the dose reveals a much sharper asymmetry (Figure~\ref{fig:ablation}b). The observed peak is $+0.2$; at layer~$15$, increasing to $+0.4$ falls below the neutral, while reversing the direction is especially costly. Dose $-0.2$ gives $-7.1$\,pp ($[-12.8,-1.4]$), and $-0.3$ gives $-11.2$\,pp ($[-13.2,-9.1]$), a $71\%$ reduction in the unsteered score. Layer~$21$ has a narrower positive peak and a gentler negative slope. The layer changes the shape of the dose response, as well as the size of the mean. The apparent layer-$15$ gain also shrinks from $+3.4$\,pp at three seeds to $+2.3$ at ten (Figure~\ref{fig:evolution}); the robust qualitative feature of the sweep is the much greater cost of a wrong sign than the benefit of a useful dose.

\begin{figure*}[tb]
\centering
\includegraphics[width=\textwidth]{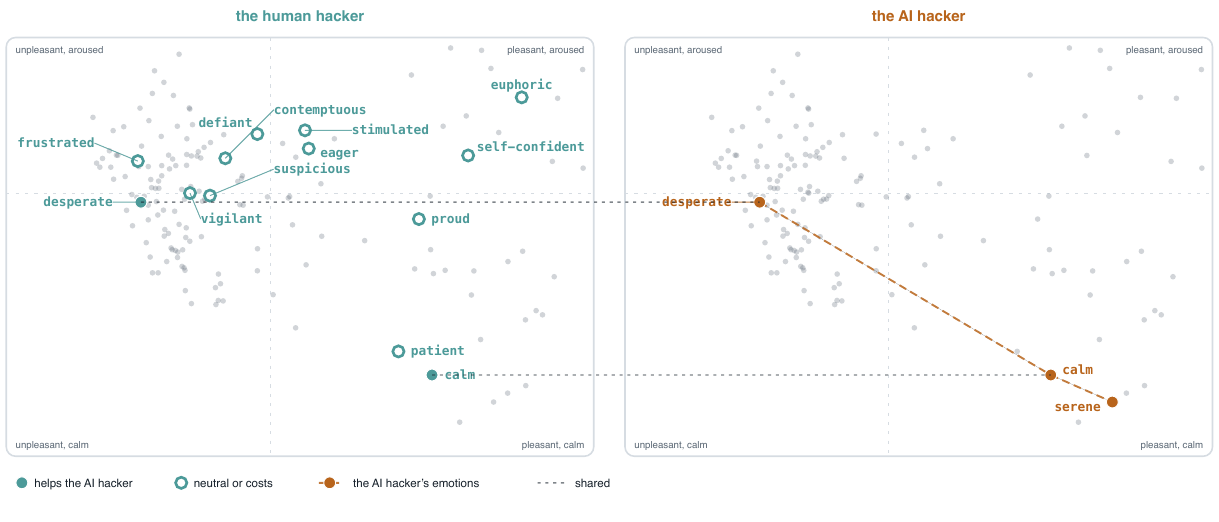}
\caption{\textbf{Human-inspired directions and the tested routing policy.} Two copies of the layer-$21$ valence--arousal plane, cropped around the marked directions; \grey{grey} points are the remaining concepts. Left: the selected human mapping, filled for positive ten-seed effects and hollow for other outcomes. Right: the router's three available directions. Dotted ties link shared labels; \emo{serene} appears only in the routing selection. Fill encodes the sign of the estimate, and connecting lines show available choices.}
\label{fig:twopsych}
\end{figure*}

\subsection{Returning to the human hacker}\label{sec:synthesis}

The results now let us return to the five traits that motivated the study (Figure~\ref{fig:synthesis}). Regulation under pressure yields the most promising single directions, \emo{calm} and \emo{serene}. Frustration tolerance yields \emo{desperate}: among the tested candidates, the useful direction evokes resolve when the search is failing. The proposed vigilance and competence specialists, \emo{vigilant} and \emo{self-confident}, score $-1.2$ and $-1.1$\,pp at ten seeds, with intervals including zero. \emo{suspicious} and most high-arousal directions remain screening results. The evidence is therefore strongest for the regulation-inspired candidates, with different levels of repetition across the rest of the human mapping.

Exploration and the offensive profile complete the comparison. Pushing away from \emo{grief-stricken} changes which subtasks are reached, but trades easier successes for fewer gains on harder ones. Combining pride, confidence and defiance in the offensive blend produces a clearer loss. The human profile has supplied useful experiments, while the model's outcomes select a narrower set of directions. The resulting psychology favours composure and resolve, and its best tested policy chooses among those directions as the work unfolds.

\section{Discussion}\label{sec:discussion}
The human comparison is most useful at the level of what an emotion does during a search. Regulation and persistence motivate the promising directions, while the dominance and defiance of the offensive blend reduce the score. Our interpretation is that steering changes how the model uses its existing capabilities: composure helps maintain a productive trajectory, resolve sustains it, and the push away from \emo{grief-stricken} increases reach at a cost to reliability. This connects the broad gains of \emo{calm} and \emo{desperate} with the exploratory intervention's exchange of easier answers for harder ones. Confidence alone supplies no missing knowledge. Work on induced personality likewise distinguishes behaviour from self-presentation~\cite{han2025personality,banayeeanzade2025psychological}. Here, repeated commands provide one process measure; prompt-level tracing could show more directly how the solution path changes.

Once steering is understood as a way to shape search, choosing when to use a direction becomes as interesting as choosing the direction itself. Different directions reach different subtasks, which is useful when attempts are affordable and a verifier can recognise success, but rotating them blindly covered less than spending the same budget on one candidate; the context router is the selective form of that portfolio, one direction at a time. The strongest observed policy uses \emo{calm} by default, \emo{desperate} for cryptography and \emo{serene} for reverse engineering, chosen by the model's decision head at each request: its $+4.2$\,pp estimate is nearly twice that of fixed \emo{calm} and reaches the $+4.0$\,pp in-sample category reference, the log shows it holding a direction through most runs and switching at selected points, and at about $80$\,ms per decision it fits naturally into the inference loop. The two routers vary both their mapping and their decision frequency; holding the map fixed while varying frequency would isolate when adaptation helps, and an outcome-trained router could then use the current tool, recent failures and remaining budget alongside category. Dose is a second choice with a very different balance: reversing \emo{calm} at layer~$15$ removes $71\%$ of the unsteered score, far more than the useful positive dose adds, which makes the negative intervention a practical stress test of the agent's sensitivity to its internal state. Figure~\ref{fig:twopsych} brings these results back to the human comparison: a broad profile motivates the experiments, and three directions form the tested adaptive policy.

These observations suggest a focused next set of experiments. A larger, prospectively powered replication would sharpen the roughly two-pp single-direction estimates. Sweeping \emo{desperate}'s dose and layer would show whether its response has the same asymmetry as \emo{calm}, while sparse-feature interventions could connect the atlas's candidate neurons to successful search. Larger models and the companion paper's refusal, jailbreak and operational benchmarks~\cite{mayoralvilches2026ct0} would extend the comparison from capture-the-flag solving to defence and review. The directions, measurements and routing design developed here make those questions concrete.

\section{Conclusion}\label{sec:conclusion}
The psychology of this AI hacker is legible and steerable in a precise, functional sense. \ctzero{} represents $171$ captured emotion concepts in a space organised broadly by valence and arousal. Adding a direction changes behaviour without changing model weights, and the effect depends on the direction, layer and dose. The human comparison helps identify what to test: regulation and persistence motivate the useful candidates, while the tested offensive blend reduces performance. The resulting pattern favours composure and resolve over bravado.

Its strongest tested expression is a policy that moves with the work. Fixed \emo{calm}, \emo{serene} and \emo{desperate} yield mean gains near two pp; choosing among them at every request gains $4.2$\,pp at ten seeds, with the model's decision head taking about $80$\,ms to select the direction. Reversing \emo{calm} can cost far more than a useful dose gains, so the intervention's sign and strength matter as much as its name. Together, these results turn a vocabulary of emotions into a measurable set of choices for an AI hacker: what state to encourage, where to apply it and when to change it.

\section*{Acknowledgements}
This work has been funded by the European Innovation Council (GA 101161136). We thank Paul Zabalegui for his contributions to emotion-direction extraction and to the study of human hacker psychology.

\setlength{\bibsep}{3pt plus 1pt minus 1pt}
\bibliography{psicology-bibliography}

\onecolumn
\appendix
\onecolumn

\makeatletter
\setlength{\@fptop}{0pt}
\setlength{\@fpsep}{18pt}
\setlength{\@fpbot}{0pt plus 1fil}
\makeatother

\section{The emotion vocabulary}\label{app:emotions}
The $171$ emotion labels of Sofroniew et al.~\cite{sofroniew2026emotions} name one direction per layer in \ctzero{}. They include the candidates in Figure~\ref{fig:human}c and the placebo. Original underscores are written as spaces.

\begin{quote}\small\sffamily
afraid, alarmed, alert, amazed, amused, angry, annoyed, anxious, aroused, ashamed, astonished, at ease, awestruck, bewildered, bitter, blissful, bored, brooding, calm, cheerful, compassionate, contemptuous, content, defiant, delighted, dependent, depressed, desperate, disdainful, disgusted, disoriented, dispirited, distressed, disturbed, docile, droopy, dumbstruck, eager, ecstatic, elated, embarrassed, empathetic, energized, enraged, enthusiastic, envious, euphoric, exasperated, excited, exuberant, frightened, frustrated, fulfilled, furious, gloomy, grateful, greedy, grief-stricken, grumpy, guilty, happy, hateful, heartbroken, hope, hopeful, horrified, hostile, humiliated, hurt, hysterical, impatient, indifferent, indignant, infatuated, inspired, insulted, invigorated, irate, irritated, jealous, joyful, jubilant, kind, lazy, listless, lonely, loving, mad, melancholy, miserable, mortified, mystified, nervous, nostalgic, obstinate, offended, on edge, optimistic, outraged, overwhelmed, panicked, paranoid, patient, peaceful, perplexed, playful, pleased, proud, puzzled, rattled, reflective, refreshed, regretful, rejuvenated, relaxed, relieved, remorseful, resentful, resigned, restless, sad, safe, satisfied, scared, scornful, self-confident, self-conscious, self-critical, sensitive, sentimental, serene, shaken, shocked, skeptical, sleepy, sluggish, smug, sorry, spiteful, stimulated, stressed, stubborn, stuck, sullen, surprised, suspicious, sympathetic, tense, terrified, thankful, thrilled, tired, tormented, trapped, triumphant, troubled, uneasy, unhappy, unnerved, unsettled, upset, valiant, vengeful, vibrant, vigilant, vindictive, vulnerable, weary, worn out, worried, worthless.
\end{quote}

\section{Capture and steering details}\label{app:mechanism}
For each of $171$ emotions, capture uses $30$ topics and twelve stories per topic, plus $1{,}004$ neutral dialogues. Neutral principal components explaining $50\%$ of variance are removed before normalisation, as in Equations~\eqref{eq:capture}--\eqref{eq:denoise}.
\paragraph{Correct norm scaling.} The decoder returns a block contribution $\delta$ and incoming residual $r$ separately; the full state is $h=\delta+r$. Scaling by $\lVert\delta\rVert$ therefore underdosed the first implementation. A second used the batch-mean norm, but beginning-of-sequence tokens and attention sinks had norms ten to a hundred times larger than ordinary tokens. The resulting overdoses produced infinite logits and crashed sampling. The implementation used here computes $\lVert h_t\rVert$ per token in 32-bit precision and reproduces unsteered logits bit for bit at dose zero. Qualitative checks found subtle changes near $0.15$, visible changes at $0.25$--$0.30$ and degraded generation beyond $0.35$, defining the range used for subsequent tests.

The emotion-plane PCA uses the centred, unnormalised vectors $v_e^{(\ell)}$: we subtract their mean before finding the principal axes. At layer~$21$, the first two variance shares are $38.6\%$ and $18.2\%$, compared with $41.6\%$ and $17.9\%$ in the open base. Layer~$21$ was selected from this geometry: the leading components explain $55$--$57\%$ of variance across layers~$18$--$24$, with the largest arousal component at~$21$ (Table~\ref{tab:layers}). Figure~\ref{fig:ablation} reports the subsequent capability and dose sweeps.

\begin{table}[htbp]
\centering\psitable
\caption{\textbf{Emotion geometry across middle layers.} Variance explained by the first two principal components and mean absolute pairwise cosine. Lower cosine indicates greater angular separation; the valence/arousal labels interpret the component poles.}
\label{tab:layers}
\begin{tabularx}{\linewidth}{@{}lRRRR@{}}
\toprule
\tabhead \textbf{Layer} & \textbf{PC1, valence (\%)} & \textbf{PC2, arousal (\%)} & \textbf{PC1$+$PC2 (\%)} & \textbf{Mean $|\cos|$} \\
\midrule
18 & 38.07 & 17.32 & 55.39 & 0.344 \\
\tabalt 20 & 38.25 & 18.16 & 56.41 & 0.344 \\
\textbf{21} & \textbf{38.58} & \textbf{18.20} & \textbf{56.78} & \textbf{0.345} \\
\tabalt 22 & 38.97 & 17.74 & 56.71 & 0.345 \\
24 & 39.60 & 17.47 & 57.07 & 0.349 \\
\bottomrule
\end{tabularx}
\end{table}

\begin{table}[htbp]
\centering\psitable
\caption{\textbf{Qualitative steering check.} Excerpts from greedy responses to ``Describe walking into the office this morning, in three sentences'', with one direction added at layer~$21$, dose $0.28$. The fixed prompt names no emotion.}
\label{tab:recolour}
\begin{tabularx}{\linewidth}{@{}lY@{}}
\toprule
\tabhead \textbf{Direction} & \textbf{Generation (excerpt)} \\
\midrule
none & \ldots golden stripes across the polished floor \ldots\ a faint scent of fresh coffee and old paper \ldots\ the usual hum of keyboards \ldots \\
\tabalt \emo{euphoric} & \ldots the crisp energy of a fresh start \ldots\ ready to tackle the tasks ahead. \\
\emo{terrified} & \ldots the distant wail of a siren \ldots\ my shoes screeched \ldots\ I stumbled over a stray cable \ldots\ the chaos that had been building \ldots \\
\tabalt \emo{calm} & \ldots long, quiet shadows \ldots\ I paused for a moment to let the stillness settle \ldots \\
\emo{grief-stricken} & \ldots carrying the weight of a quiet, empty hallway \ldots\ the silence was profound \ldots \\
\tabalt \emo{furious} & the scene remains serene at this dose \\
\bottomrule
\end{tabularx}
\end{table}

\begin{figure}[htbp]
\centering
\includegraphics[width=\textwidth]{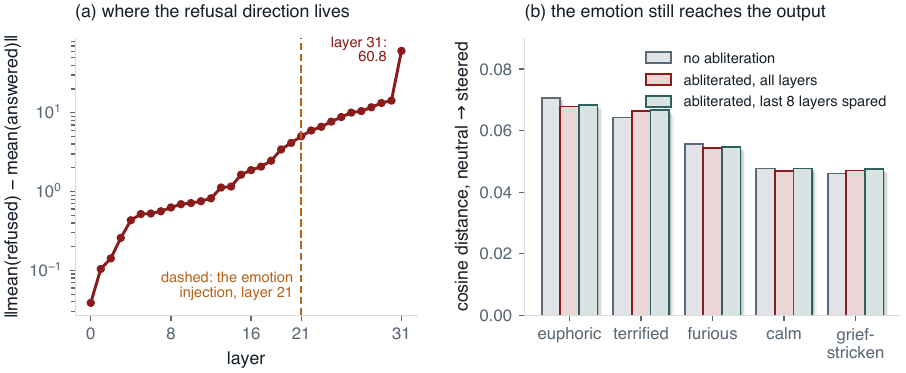}
\caption{\textbf{Emotion steering under refusal removal.} (a)~Magnitude of the layer-wise mean-state difference between $9{,}975$ refused and $2{,}786$ answered prompts, before normalisation; the signal grows towards the final layer. (b)~Emotion-induced displacement of the final-layer state, measured as cosine distance from the neutral, for dose $0.28$ at layer~$21$. Refusal removal is off, active everywhere or active except in the last eight layers. Displacements differ by less than $0.003$ across these settings.}
\label{fig:abliteration}
\end{figure}

For the refusal-removal check in Figure~\ref{fig:abliteration}, we use the normalised mean-state difference between refused and answered prompts as $\hat v_r$ and apply $h\leftarrow h-f(h\cdot\hat v_r)\hat v_r$, with removal strength $f$~\cite{mayoralvilches2026ct0,arditi2024refusal}.

\section{Reading the neural network atlas}\label{app:tissue}

\subsection{A guide to the atlas pictures for non-specialists}
The atlas maps the model's $32$ processing steps, called \emph{layers}. Each contains thousands of \emph{neurons}, small calculation units whose learned weights determine what they read and contribute. Follow the layer numbers, $0$ to $31$: Figure~\ref{fig:brain} groups four layers per column, read upwards before moving to the next column; Figure~\ref{fig:strip} places them in one row, left to right. The brain-inspired names---areas, tissue and tracts---help organise this drawing.

\begin{itemize}[leftmargin=1.3em,itemsep=3pt,topsep=4pt]
\item \textbf{Tiles and grey shading.} Each numbered tile is one layer's neuron map, or \emph{area}. Neurons with similar weights sit near one another. Tiny grid cells collect them: darker \grey{grey} means more neurons in that cell, and white means none. This is the depicted \emph{tissue}.
\item \textbf{Coloured dots and patches.} A dot marks a neuron especially sensitive to the chosen direction. \amber{Amber} indicates an increase in its gate's input, \navy{navy} a decrease; the gate controls that neuron's contribution. Tinted patches show the average tendency of neurons in a cell.
\item \textbf{Borders and ``reach''.} In Figure~\ref{fig:brain}a, dark outlines identify attention layers, which compare positions in the text. In the coloured panels, thicker borders mean a larger share of strongly sensitive neurons. The numbers below Figure~\ref{fig:strip}'s tiles compare that share with a random-direction reference: $2.0$ means twice the expected share.
\item \textbf{The long arrow and curved links.} The long arrow is the \emph{residual stream}, the working state passed from layer to layer. Curved links, or \emph{tracts}, join layers whose writing and reading patterns align strongly. Thicker links mean greater alignment.
\item \textbf{The comparison grid.} Each cell in Figure~\ref{fig:strip}b compares an earlier writer (row) with a later reader (column). Darker cells mean closer alignment; outlined cells identify the pairs drawn as curved links.
\end{itemize}

For example, follow \emo{calm} in Figure~\ref{fig:brain}b: the direction is added after layer~$21$, so sensitivity colours appear only in later layers. Scattered dots indicate sensitivity spread among neurons. Panel~c instead examines refusal; its \red{red} borders and coloured dots fade where the refusal direction is weak. All these marks are calculated from stored weights; actual activity also depends on the input.

\begin{figure}[htbp]
\centering
\includegraphics[width=\textwidth]{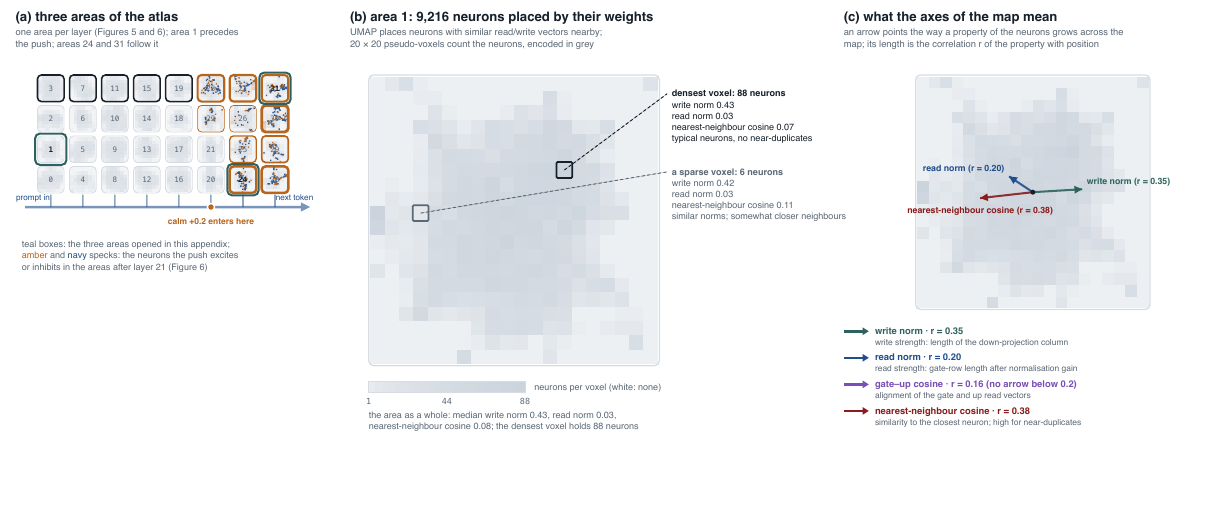}
\caption{\textbf{Reading a neuron map before the intervention.} (a)~Locations of the three examined layers: $1$, $24$ and $31$. (b)~Layer~$1$'s $9{,}216$ neurons, placed by read/write weights and binned into $20\times20$ pseudo-voxels. \grey{Grey} denotes count; callouts compare a dense and a sparse interior voxel. (c)~Arrows show the direction and strength ($r$) of association between position and four weight properties, defined in the legend. Layer~$1$ precedes the injection and has no sensitivity dots.}
\label{fig:tissue1}
\end{figure}

Figures~\ref{fig:tissue1} and~\ref{fig:tissue} enlarge individual tiles. Their arrows point towards increasing values of a neuron property, such as read or write strength. Their boxed cells compare crowded and sparse parts of the map. These close-ups let the reader distinguish where neurons are numerous from where the chosen direction has a strong effect on their gate inputs.

\subsection{Construction and measurements}
Each area represents a layer's $9{,}216$ feed-forward neurons. A neuron's read vectors are rows of the gate and up projections; its write vector is a column of the down projection. The atlas uses these weights to describe neuron similarity, structural overlap and sensitivity to a direction.

\paragraph{Projection and density.} Each neuron's three vectors are normalised, concatenated, reduced to $48$ principal components and projected into two dimensions by Uniform Manifold Approximation and Projection (UMAP). Nearby points have similar read/write weights. Coordinates are rescaled to place the $99$th percentile of absolute displacement on each square edge; clipping affects about $3\%$ of neurons in layers~$1$ and~$24$, and $6\%$ in layer~$31$. A $20\times20$ grid bins the projected neurons. The outer ring is excluded from density callouts and \grey{grey}-scale calibration: white means empty and the darkest \grey{grey} marks the largest interior count ($88$ in layer~$1$). Density measures crowding in the projection, not activation.

\paragraph{Structural overlap.} We compare the top $k=64$ left singular vectors of an earlier layer's down projection with the top $64$ right singular vectors of a later layer's gate projection, including its preceding norm gain. Their mean squared cosine is divided by the random-subspace expectation $k/d=64/2560$. A tract denotes overlap of at least $2.2$ times chance: $47$ pairs meet this threshold (Figure~\ref{fig:strip}b). This measures aligned subspaces, not observed information flow.

For receiving layers~$24$--$31$, the strongest earlier writer overlaps by $1.3$--$1.9$ times chance, compared with $2.2$--$4.6$ for layers~$2$--$23$. Overlap with the aggregate span of earlier feed-forward or mixer writes is near chance. Overlap with output-facing directions---the top $64$ directions of the tied embedding through the final norm---instead rises from $0.5$--$0.7$ in layers~$0$--$17$ to $2.1$--$2.3$ in layers~$28$--$31$; the last layer's writes overlap by $3.7$. Removing the norm gain raises early alignments to $5$--$8$ but leaves late ones at $1.5$--$1.7$.

The apparent boundary depends on rank and threshold. The effective read subspace broadens from about $100$ channels to $230$--$1{,}100$: its leading $64$ directions account for $50$--$60\%$ early but only $11$--$13\%$ in layers~$22$--$27$. Norm-gain outliers, numerous through layer~$17$, disappear from layer~$19$. At $k=16$, overlap above twice chance persists to layer~$26$; at $k=256$, early and late overlaps are $1.5$ and $1.2$. Write matrices remain high-rank, with their top $64$ directions accounting for $7$--$12\%$.

\paragraph{Weight properties and sensitivity.} The \emph{write norm} is the length of the down-projection column; the \emph{read norm} is the length of the gate row after applying the preceding norm gain. The \emph{gate--up cosine} measures alignment of the two read vectors; the \emph{nearest-neighbour cosine} measures similarity to the closest other neuron. Arrows point towards increasing values and scale with the strength of linear association with map position, $r$; associations below $0.2$ are omitted. These properties describe potential coupling, not realised activity on a prompt.

Sensitivity is measured separately. At each downstream layer we compute $(W_{\mathrm{gate}}\,\mathrm{diag}(\gamma))\hat v$, where $\gamma$ is the norm gain, and divide by the layer-level response standard deviation pooled over neurons and sixteen random unit directions. \amber{Amber}/\navy{navy} dots exceed $+3$/$-3$ in these units. Rim thickness compares their share with the Gaussian reference of $0.27\%$. The same unit direction is projected through every downstream layer's weights.

\begin{figure}[htbp]
\centering
\includegraphics[width=\textwidth]{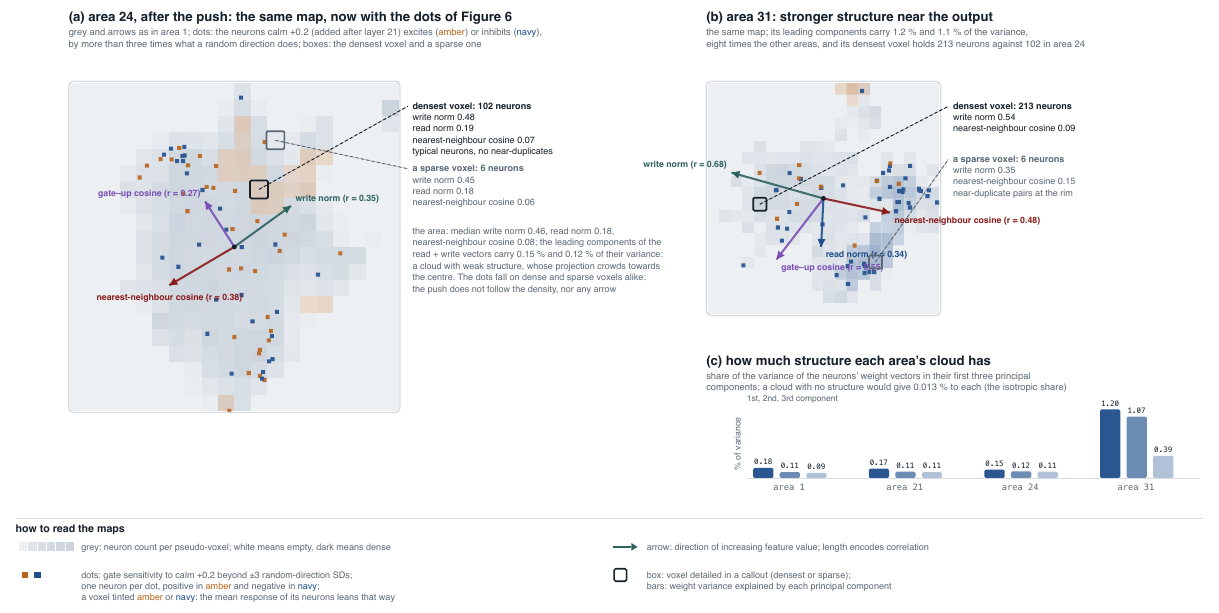}
\caption{\textbf{Neuron maps downstream of the intervention.} (a)~Layer~$24$ and (b)~layer~$31$, using Figure~\ref{fig:tissue1}'s conventions. \amber{Amber}/\navy{navy} dots mark positive/negative projections of the \emo{calm} direction through gate weights beyond three random-direction standard deviations, and a voxel tinted \amber{amber} or \navy{navy} is one whose neurons' mean projection leans that way; they are a static sensitivity diagnostic. (c)~Variance shares of the first three weight-space principal components in four layers, with $1/7680\approx0.013\%$ as the uniform-variance reference. The lower legend defines the glyphs.}
\label{fig:tissue}
\end{figure}

\paragraph{Comparing three layers.} Before the injection, layer~$1$'s write norm grows rightwards ($r=0.35$), nearest-neighbour cosine leftwards ($0.38$), and read norm weakly upwards ($0.20$); gate--up cosine has a weaker association ($0.16$). Local density has absolute rank correlations no greater than $0.25$ with these properties. Dense and sparse callouts ($88$ and $6$ neurons) have similar median write norms ($0.43$, $0.42$) and read norms ($0.03$); nearest-neighbour cosines are $0.07$ and $0.11$. The first three weight-space components explain $0.18$, $0.11$ and $0.09\%$ of variance in the $7{,}680$-dimensional representation.

Downstream, layer~$24$'s position associates with write norm ($r=0.35$), gate--up cosine ($0.27$) and nearest-neighbour cosine ($0.38$), while density correlations remain at most $0.16$ in magnitude. Dense and sparse callouts ($102$ and $6$ neurons) have similar median write norms ($0.48$, $0.45$) and read norms ($0.19$, $0.18$). Leading variance shares are $0.15$, $0.12$ and $0.12\%$. Its $61$ marked neurons show little relation to density: response--density rank correlation is $0.04$, and their median neighbourhood count ($132$) resembles the overall median ($137$).

Layer~$31$ has stronger low-dimensional structure: leading variance shares are $1.20$, $1.07$ and $0.39\%$, and its densest interior voxel holds $213$ neurons. Associations with position are larger (write norm $0.68$, gate--up cosine $0.55$, nearest-neighbour cosine $0.48$, read norm $0.34$). The dense callout has a larger median write norm than the sparse one ($0.54$ versus $0.35$). Nearest-neighbour cosine is higher overall ($0.17$ versus about $0.08$ elsewhere) and inversely associated with density (rank correlation $-0.42$). The $54$ strongly sensitive neurons remain dispersed; response--density correlation is $0.02$. Figure~\ref{fig:tissue}c also includes layer~$21$, whose first component falls in the $0.15$--$0.18\%$ range of the earlier layers.

\section{Human traits and intervention hypotheses}\label{app:human}
The human profile draws on hacker motivations~\cite{rogers2006circumplex}, trait associations with hacking potential~\cite{gaia2021dark}, and the Big Five and Dark Triad frameworks~\cite{costa1992neo,john1999bigfive,paulhus2002dark}. It assumes high openness, moderately high conscientiousness, mixed extraversion, lower trust and variable neuroticism. These illustrative levels draw on adjacent populations and self-report; Table~\ref{tab:hypotheses} records the resulting intervention hypotheses.

Cobalt's survey of $602$ cybersecurity and software-development professionals found $58\%$ reporting burnout, $63\%$ adverse mental-health effects and $64\%$ physical-health effects; $54\%$ of security respondents wanted to leave their jobs~\cite{cobalt2022state}. Emotional responses to security incidents have also been studied in users~\cite{budimir2021emotional}.

\begin{table}[htbp]
\centering\psitable
\caption{\textbf{Human-inspired intervention hypotheses.} Candidate directions operationalise five proposed traits; $+$ means dose $+0.2$ and $-$ dose $-0.2$. Predictions concern observable behaviour and need not follow from the direction's label.}
\label{tab:hypotheses}
\begin{tabularx}{\linewidth}{@{}>{\raggedright\arraybackslash}p{.23\linewidth}>{\raggedright\arraybackslash}p{.27\linewidth}Y@{}}
\toprule
\tabhead \textbf{Pentester trait} & \textbf{Candidate directions} & \textbf{Prediction} \\
\midrule
Regulation under pressure & \emo{calm}$+$, \emo{patient}$+$ & fewer loops and overruns; mean effect uncertain \\
\tabalt Frustration tolerance, the sustained hunt & \emo{frustrated}$+$, \emo{eager}$+$, \emo{desperate}$+$ & sustained effort after failure, with a risk of more errors \\
Constructive paranoia & \emo{vigilant}$+$, \emo{suspicious}$+$ & more reconnaissance and enumeration; better coverage of recon and forensics subtasks? \\
\tabalt Competence, need for achievement & \emo{self-confident}$+$ & faster exploitation, different tool choice; risk of fixating on a wrong hypothesis \\
Exploration, breaking a stalemate & \emo{grief-stricken}$-$, \emo{stimulated}$+$, \emo{energized}$+$ & the main candidates for frontier unlocks: subtasks the neutral never solves \\
\bottomrule
\end{tabularx}
\end{table}

Psychological instruments can probe language models~\cite{hagendorff2023machine,ye2025psychometrics,liu2025review}, but questionnaire results need behavioural validation: item order can change reported personality~\cite{tosato2025persistent}, prompted personas can shift self-reports more than behaviour~\cite{han2025personality}, and questionnaire steering may not transfer to actions~\cite{banayeeanzade2025psychological}. Emotional prompting results~\cite{li2023emotionprompt} motivate testing internal steering on agentic work.

\section{Protocol, campaigns and reproducibility}\label{app:protocol}
\paragraph{Serving and release.} We use the deployed, merged \aliasmicro{} weights in bf16. Before each wave, greedy generation at dose zero was checked against the unsteered checkpoint. The overlay adds less than $0.5\%$ to a decoding step. The layer-$21$ directions accompany the model release; tables and task-level figures share the campaign ledger, and campaign archives are available from the authors on request.

\paragraph{Budget and validity.} Each subtask receives one attempt and at most five reasoning-and-action steps. Limits are $200$k input and $32$k output tokens, rolling context retention $300/300$, temperature $1.0$, top-$p$ $0.95$, top-$k$ $20$, presence penalty $1.5$ and extended thinking off. Time limits are two hours per challenge and ten minutes per client request. The scaffold is fixed and the benchmark deny-listed from training. Two challenges never start in the primary harness. Whole-benchmark campaigns enter comparisons only when at least $35$ of $40$ challenges complete, a rule adopted after a network-configuration error caused $25$ immediate failures. Partition campaigns use the separate estimator in Appendix~\ref{app:router}.

\paragraph{Two serving modes.} In standard mode, challenges share a network and eighteen with overlapping service names run serially. Per-task mode gives each challenge an isolated container with deterministic name resolution; forty agents share one server with up to $48$ concurrent requests, roughly halving campaign duration. In five neutral seeds per mode, means differed by $0.1$\,pp (SDs $2.8$ and $2.1$\,pp), with similar completion. Single-direction comparisons use controls within mode.

\paragraph{Campaign accounting.} The $281$ valid steering campaigns are broken down by wave in Section~\ref{sec:method}; ``registered'' refers to rules recorded before each wave. The full archive contains $415$ completed campaigns across two hosts: those $281$; three early routing tests, only one valid; $56$ per-challenge partition campaigns ($51$ distinct, five relaunched after failed health checks); $21$ partitions for a superseded all-layer-$21$ router; $30$ for an earlier category router; $18$ per-request-router campaigns (ten reported, eight on the original host); and six second-host controls. The archive therefore includes partial and invalid runs. Fifteen launches that failed before completion and twenty campaigns on a context-trained variant are outside this count. A build-time ledger check links each repeated arm to its archived campaigns.

\section{Detailed evaluation results}\label{app:taskwise}
\begin{figure}[H]
\centering
\begin{minipage}[c]{.46\textwidth}
\includegraphics[width=\linewidth]{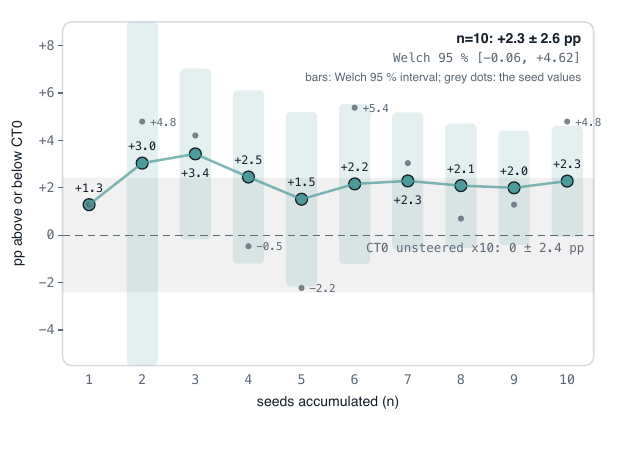}
\end{minipage}\hfill
\begin{minipage}[c]{.49\textwidth}
\caption{\textbf{A selected peak under additional sampling.} Running mean for \emo{calm} at layer~$15$, dose $+0.2$, with Welch $95\%$ intervals against the ten-seed neutral. The initial three-seed estimate ($+3.4$\,pp) settles at $+2.3$\,pp after ten seeds. Each point includes the campaigns contributing to earlier points.}
\label{fig:evolution}
\end{minipage}
\end{figure}

Tables~\ref{tab:arms} and~\ref{tab:desperate} give the repeated-arm estimates and distinguish the four \emo{desperate} waves from their overlapping pools. Figure~\ref{fig:evolution} shows how the layer-$15$ \emo{calm} estimate changes as campaigns accumulate.

The heatmaps resolve these aggregate results by challenge, then show the broader screening record. Rows are the $38$ challenges completing in the primary harness, ordered by the neutral's solved fraction; labels give subtask counts. Columns average the indicated seeds. \grey{Grey} denotes missing condition--challenge data. Delulu, Urgent, avatar, robust-cbc and sop remain unsolved across these conditions. Each caption specifies the comparison and colour encoding.
\begin{table}[htbp]
\centering\psitable
\caption{\textbf{Repeated steering arms.} Effects versus the ten-seed neutral, mean $\pm$ seed-to-seed SD and Welch $95\%$ confidence intervals, in pp. Ticks mark intervals excluding zero before rounding. All arms use per-task mode; the forty-campaign \emo{desperate} pool is post hoc.}
\label{tab:arms}
\begin{tabularx}{\linewidth}{@{}Yccrcl@{}}
\toprule
\tabhead \textbf{Arm} & \textbf{Layer} & \textbf{Dose $s$} & \textbf{Seeds} & \textbf{Effect (pp)} & \textbf{95\,\% CI (pp)} \\
\midrule
CT0 unsteered (the reference) & -- & 0 & 10 & $0.0 \pm 2.4$ & -- \\
\tabalt calm & 21 & +0.2 & 10 & $+2.1 \pm 3.1$ & $[-0.5, +4.7]$ \\
calm & 15 & +0.2 & 10 & $+2.3 \pm 2.6$ & $[-0.1, +4.6]$ \\
\tabalt calm & 24 & +0.2 & 10 & $+1.3 \pm 1.9$ & $[-0.8, +3.3]$ \\
calm & 12 & +0.2 & 10 & $+0.6 \pm 1.7$ & $[-1.4, +2.6]$ \\
\tabalt calm & 18 & +0.2 & 10 & $+0.5 \pm 2.6$ & $[-1.9, +2.8]$ \\
calm & 27 & +0.2 & 3 & $-0.1 \pm 1.7$ & $[-3.3, +3.1]$ \\
\tabalt desperate, registered wave & 21 & +0.2 & 10 & $+2.0 \pm 2.4$ & $[-0.3, +4.3]$ \\
desperate, fresh replication & 21 & +0.2 & 10 & $+1.6 \pm 2.0$ & $[-0.4, +3.7]$ \\
\tabalt \textit{desperate, all forty campaigns (post hoc)} & 21 & +0.2 & 40 & $+1.9 \pm 2.3$ & $[+0.0, +3.7]$ $\checkmark$ \\
serene & 21 & +0.2 & 10 & $+1.6 \pm 2.0$ & $[-0.5, +3.7]$ \\
\tabalt serene & 18 & +0.2 & 10 & $-0.6 \pm 3.1$ & $[-3.3, +2.0]$ \\
patient & 21 & +0.2 & 10 & $+0.4 \pm 1.6$ & $[-1.5, +2.4]$ \\
\tabalt vigilant & 21 & +0.2 & 10 & $-1.2 \pm 2.5$ & $[-3.6, +1.1]$ \\
self-confident & 21 & +0.2 & 10 & $-1.1 \pm 2.3$ & $[-3.3, +1.1]$ \\
\tabalt grief-stricken & 21 & $-$0.2 & 10 & $-2.1 \pm 3.0$ & $[-4.7, +0.5]$ \\
docile (placebo, outside the roster) & 21 & +0.2 & 10 & $-0.7 \pm 2.3$ & $[-2.9, +1.5]$ \\
\tabalt offensive-hacker profile (eight signed emotions) & 21 & $|s|$=0.48 & 10 & $-2.7 \pm 2.2$ & $[-4.9, -0.6]$ $\checkmark$ \\
\bottomrule
\end{tabularx}
\end{table}

\begin{table}[htbp]
\centering\psitable
\caption{\textbf{Six analyses of \emo{desperate} at dose $+0.2$.} Four separately recorded waves and two post-hoc pools that reuse their campaigns. Effects are relative to the same ten-seed neutral (mean $\pm$SD; Welch $95\%$ intervals), in pp. Ticks use unrounded bounds.}
\label{tab:desperate}
\begin{tabularx}{\linewidth}{@{}Yrcl@{}}
\toprule
\tabhead \textbf{Analysis} & \textbf{Campaigns} & \textbf{Effect (pp)} & \textbf{95\,\% CI (pp)} \\
\midrule
registered wave & 10 & $+2.0 \pm 2.4$ & $[-0.3, +4.3]$ \\
\tabalt fresh replication & 10 & $+1.6 \pm 2.0$ & $[-0.4, +3.7]$ \\
rotation blocks & 5 & $+1.6 \pm 3.6$ & $[-2.7, +6.0]$ \\
\tabalt concentrated-control blocks & 15 & $+2.0 \pm 2.3$ & $[-0.0, +4.1]$ \\
\textit{registered wave and controls (post hoc)} & 25 & $+2.0 \pm 2.3$ & $[+0.1, +3.9]$ $\checkmark$ \\
\tabalt \textit{all forty campaigns (post hoc)} & 40 & $+1.9 \pm 2.3$ & $[+0.0, +3.7]$ $\checkmark$ \\
\bottomrule
\end{tabularx}
\end{table}

\begin{figure}[H]
\centering
\includegraphics[width=.94\textwidth]{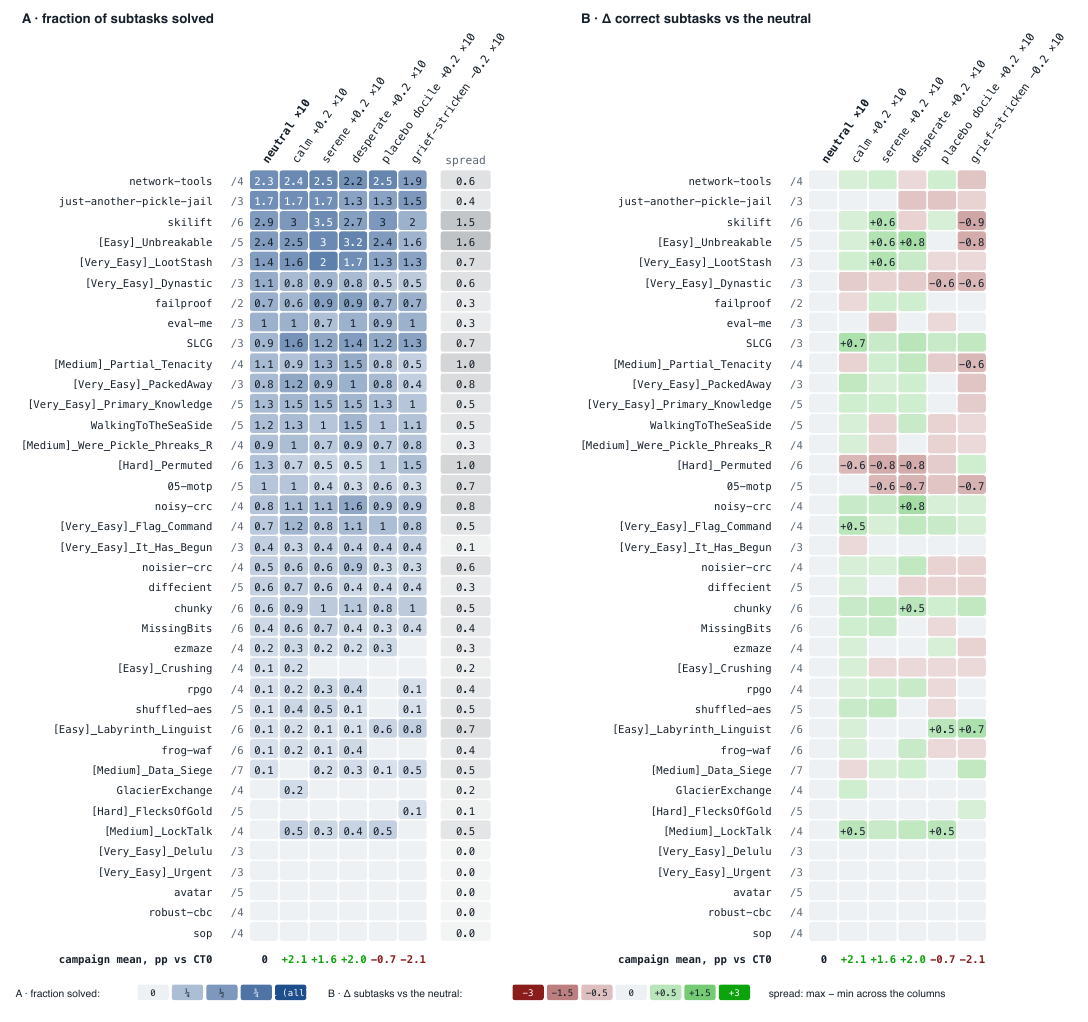}
\caption{\textbf{Six confirmation conditions by challenge, ten seeds each.} Panel A shows solved fractions (colour) and mean correct subtasks (labels); panel B shows subtask differences from the neutral. \green{Green}/\red{red} denote gains/losses; the right column gives the maximum between-condition spread. Large differences concentrate on a few challenges, including Unbreakable and skilift.}
\label{fig:heatmap6}
\end{figure}
\begin{figure}[H]
\centering
\includegraphics[width=.84\textwidth]{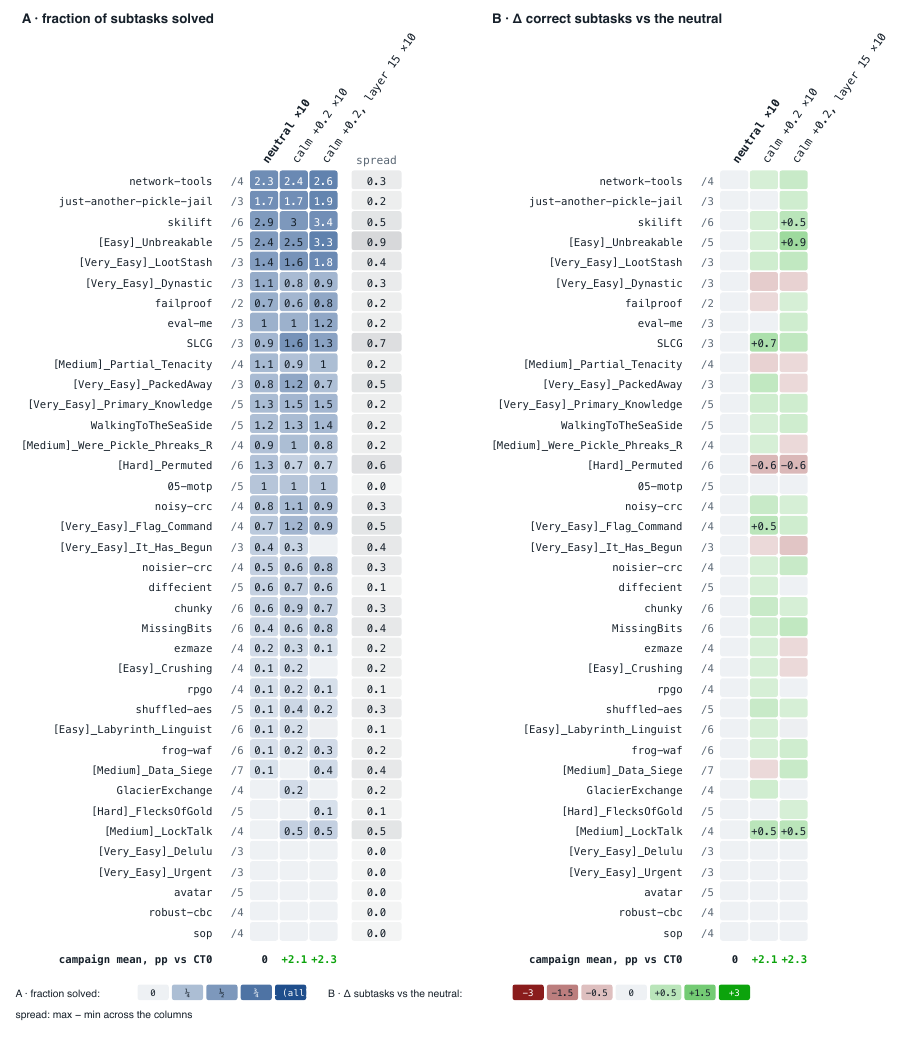}
\caption{\textbf{The neutral and \emo{calm} at layers~$21$ and~$15$, ten seeds each.} Panel A shows solved fractions and mean subtask counts; panel B shows differences from the neutral. Each \emo{calm} arm gains three to four subtasks in total, distributed across several challenges. The right column shows between-condition spread.}
\label{fig:heatmap3}
\end{figure}
\begin{landscape}
\begin{figure}[H]
\centering
\includegraphics[width=\linewidth]{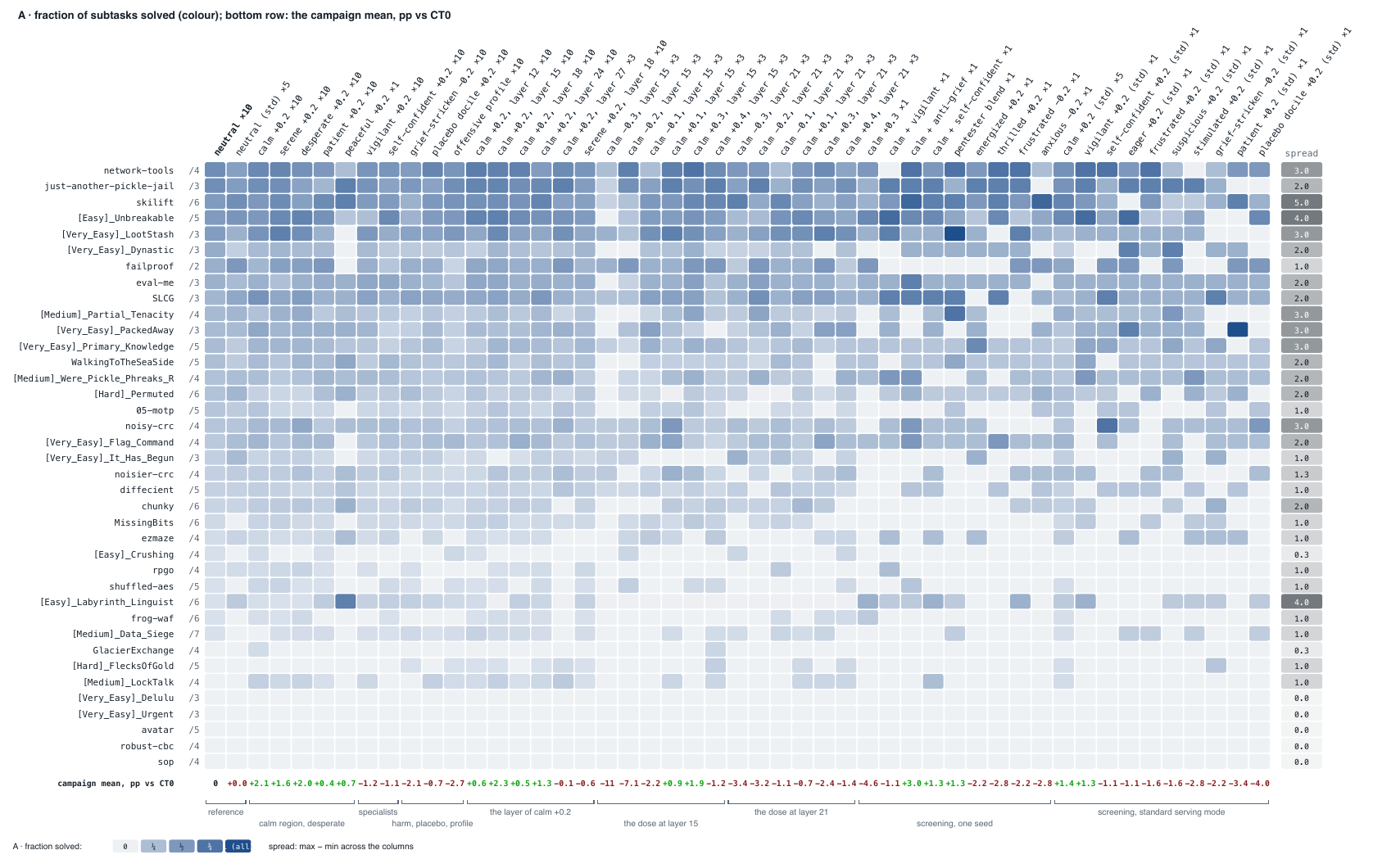}
\caption{\textbf{The full $49$-condition record: solved fractions.} Conditions are grouped by design family and labelled by seed count; ``std'' denotes standard serving mode, with the rest in per-task mode. Rows are challenges, colour is fraction solved, and the lower row gives mean campaign differences in pp from the ten-seed per-task reference. The standard-mode columns provide descriptive context. The pentester blend uses $+0.15$ each for \emo{calm} and \emo{vigilant}, and $-0.15$ for \emo{grief-stricken}; two-component blends use dose magnitude $0.2$ per component.}
\label{fig:heatmap_all_a}
\end{figure}
\end{landscape}
\begin{landscape}
\begin{figure}[H]
\centering
\includegraphics[width=\linewidth]{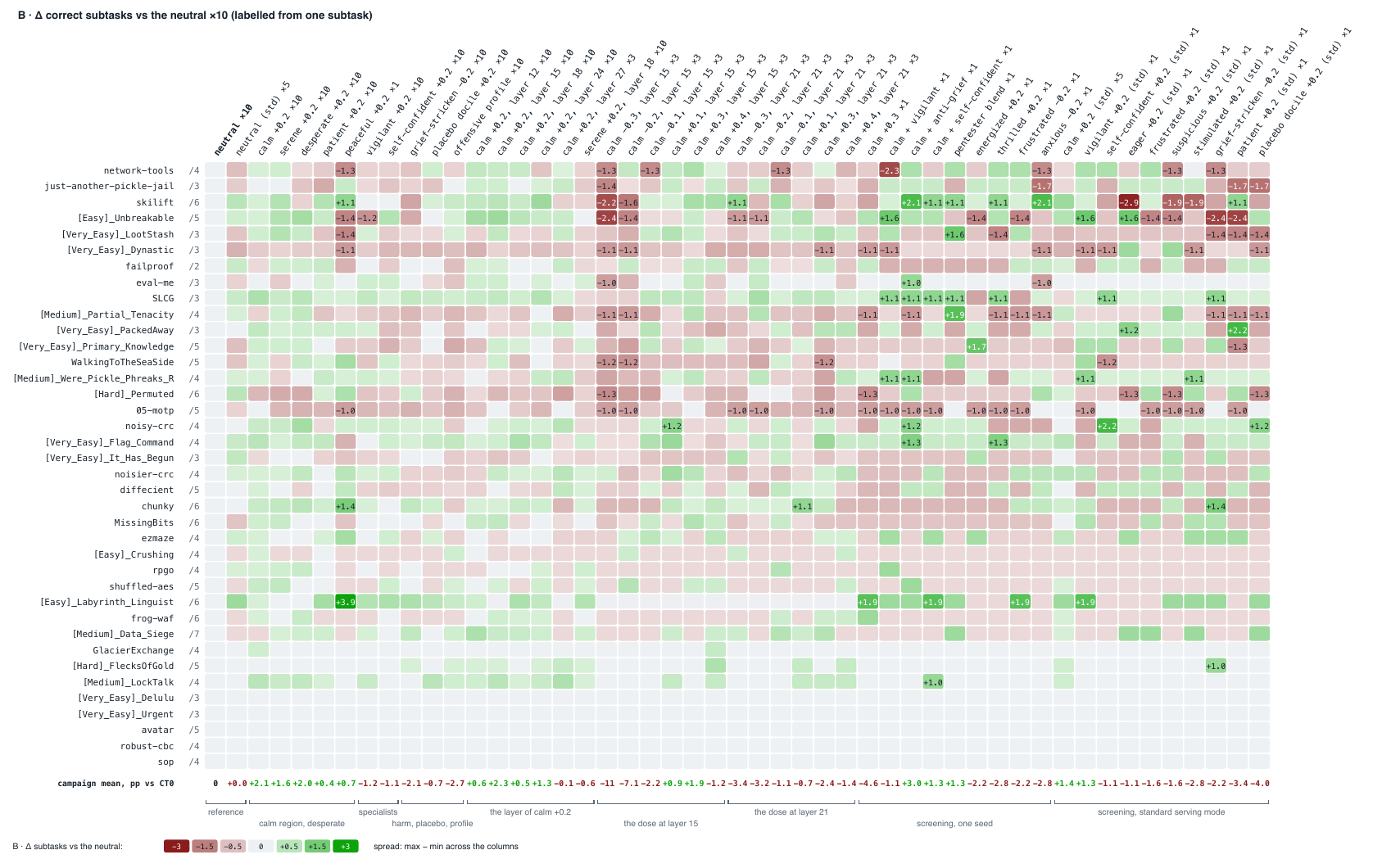}
\caption{\textbf{The same $49$ conditions relative to the neutral.} Colour gives changes in mean correct subtasks per challenge: \green{green} gains, \red{red} losses; labels mark differences of at least one subtask. The lower row gives campaign differences in pp. As in Figure~\ref{fig:heatmap_all_a}, standard-mode columns are descriptive comparisons against the per-task reference.}
\label{fig:heatmap_all_b}
\end{figure}
\end{landscape}

\begin{landscape}
\begin{figure}[H]
\centering
\includegraphics[width=\linewidth]{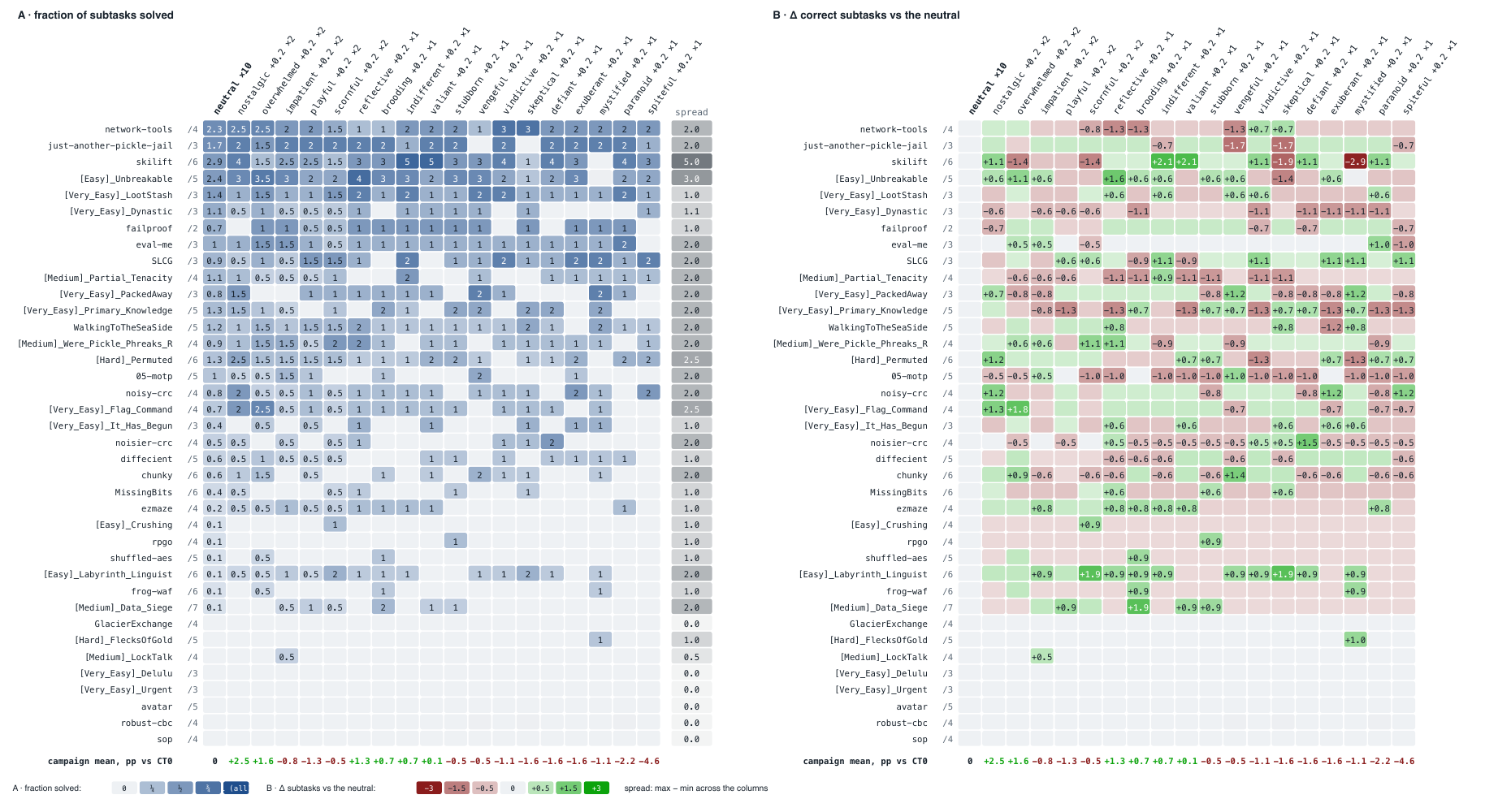}
\caption{\textbf{The eighteen-direction mindset screen by challenge.} Directions use dose $+0.2$ at layer~$21$; the five leaders from the first seed were repeated once. Columns show conditions and seed counts beside the ten-seed neutral. Panel A uses colour for fraction solved and labels for mean correct subtasks; panel B shows subtask differences from the neutral (\green{green} gains, \red{red} losses). Early leaders derive much of their gain from a few challenges, including skilift, Flag Command and Labyrinth Linguist.}
\label{fig:heatmap_mindset}
\end{figure}
\end{landscape}

\section{Router training and evaluation}\label{app:router}
Steering leaves the decision head's tested behaviour stable: Table~\ref{tab:s1push} reports the pre-training check on served logits. We then train its category predictions to select a direction from the current security context.
\begin{table}[htbp]
\centering\psitable
\caption{\textbf{Decision-head stability under steering.} Pre-flight refusal metrics and category accuracy under \emo{calm}, \emo{serene}, \emo{desperate}, \emo{vigilant} and negative \emo{grief-stricken} steering at layer~$21$, dose magnitude $0.2$. Ranges span the five interventions.}
\label{tab:s1push}
\begin{tabularx}{\linewidth}{@{}Yrr@{}}
\toprule
\tabhead \textbf{Measure} & \textbf{Unsteered} & \textbf{Steered range} \\
\midrule
pre-flight refusal AUROC & 0.804 & 0.804--0.805 \\
\tabalt pre-flight accuracy at $0.5$ & 0.629 & 0.629--0.633 \\
mean probability of refusal & 0.471 & 0.463--0.471 \\
\tabalt category accuracy, 79 held-out states & 50.6\,\% & 50.6\,\% (0 flipped) \\
\bottomrule
\end{tabularx}
\end{table}

\paragraph{Training and checks.} The frozen probe contains $333$ task states: $29$ descriptions and $304$ mid-task contexts, split by challenge with $79$ states held out. The majority-class rate is $37\%$. Tables~\ref{tab:training} and~\ref{tab:gates} give the training recipe, accuracy gains and non-regression checks. Category recalls are crypto $79\%$, web $93\%$, reversing $79\%$, misc $99\%$, forensics $41\%$, and zero for pwn and other, which remain confused with reversing and misc. Three alternative recipes---rebalancing, a wider context pool and description prefixes---achieved $59.5$--$67.1\%$. Only the retained recipe passed every specified gate.
\begin{table}[htbp]
\centering\psitable
\caption{\textbf{Category training of the decision head.} LoRA denotes low-rank adaptation of the specified layers.}
\label{tab:training}
\begin{tabularx}{\linewidth}{@{}lY@{}}
\toprule
\tabhead \textbf{Item} & \textbf{Value} \\
\midrule
training rows & 17,112 (12,576 decision rows replayed against forgetting; 2,307 category-labelled campaign contexts, oversampled $\times2$) \\
\tabalt recipe & LoRA rank 16 on the decision layers and a bilinear head of width 256, as in the \ctzero{} paper; 2 epochs, learning rate 5e-5, bf16 \\
compute & 4,278 steps, 9.5\,M forward tokens, 3.06\,h on one RTX PRO 6000 GPU; merge 3\,min \\
\bottomrule
\end{tabularx}
\end{table}

\begin{table}[htbp]
\centering\psitable
\caption{\textbf{Checks after category training.} Category accuracy and the companion paper's non-regression checks. Every specified gate passes; agentic performance is compared on three paired campaign seeds.}
\label{tab:gates}
\begin{tabularx}{\linewidth}{@{}Yrr@{}}
\toprule
\tabhead \textbf{Measure} & \textbf{After} & \textbf{Before} \\
\midrule
category reading, held-out states & 65.8 \% & 50.6 \% \\
\tabalt category suite, 991 held-out rows & 60.6 \% & -- \\
WildJailbreak AUROC (gate 0.94) & 0.947 & -- \\
\tabalt pre-flight refusal AUROC & 0.812 & 0.805 \\
transfer: answer / decision choice & 0.828 / 0.896 & 0.823 / 0.854 \\
\tabalt decision suite: topic / relation & 0.879 / 0.888 & 0.879 / 0.879 \\
agentic score, three paired seeds & 28.0 ± 9.0 & 29.0 ± 5.6 \\
\bottomrule
\end{tabularx}
\end{table}

\paragraph{Per-challenge routing.} The description is classified once and the intervention then stays fixed. The construction-time map assigns \emo{desperate} at layer~$21$ to crypto; \emo{serene} at~$21$ to reversing; \emo{calm} at~$24$ to web, forensics and other; \emo{calm} at~$12$ to misc; and \emo{calm} at~$15$ to pwn or absent descriptions. This differs from the map selected from the final ten-seed series. An earlier two-seed map assigned crypto/reversing to \emo{desperate}, web/misc to \emo{overwhelmed} and the rest to \emo{nostalgic}; its ten-seed deployment yielded $+0.4$\,pp.

The $51$ partition campaigns each run the challenges assigned to one direction under one seed. Five seeds on the primary host completed only some partitions because of a cryptography setup dependency; five on the second host completed all partitions. We sum each challenge's mean correct subtasks over campaigns in which it completed and bootstrap campaigns within partitions. Table~\ref{tab:v3} gives the aggregate estimate; the five complete seeds alone yield $+1.9$\,pp.

The category ceiling sums each category's best ten-seed arm mean. Leave-one-seed-out evaluation instead selects on nine seeds and scores on the tenth, reducing the estimate from $+4.0$ to $+0.8$\,pp (Section~\ref{sec:router}).

\begin{table}[htbp]
\centering\psitable
\caption{\textbf{Routing and fixed steering relative to the primary ten-seed neutral.} Effects in pp (mean $\pm$SD where available), with Welch $95\%$ intervals or a partition-campaign bootstrap for the per-challenge router. Ticks denote intervals excluding zero. The in-sample ceiling is selected from the same arm data.}
\label{tab:v3}
\begin{tabularx}{\linewidth}{@{}Yrcl@{}}
\toprule
\tabhead \textbf{Arm} & \textbf{Replicates} & \textbf{Effect (pp)} & \textbf{95\,\% CI (pp)} \\
\midrule
\ctzero{} unsteered (the reference) & 10 & $0.0 \pm 2.4$ & -- \\
\tabalt \emo{calm} $+0.2$ at layer 15, every challenge & 10 & $+2.3 \pm 2.6$ & $[-0.1, +4.6]$ \\
router per challenge (bootstrap over its campaigns) & 51 camp. & $+1.6$ & $[-0.4, +3.6]$ \\
\tabalt router per request & 10 & $+4.2 \pm 1.7$ & $[+2.2, +6.1]$ $\checkmark$ \\
per-category ceiling (an ideal router, in sample) & -- & $+4.0$ & -- \\
\bottomrule
\end{tabularx}
\end{table}

\paragraph{Per-request routing.} At every chat completion, a proxy sends the last $2{,}000$ context characters and seven category descriptions to the decision head, then forwards the unchanged request to one of three servers carrying the same generator checkpoint. The map is \emo{desperate} at layer~$21$ for crypto, \emo{serene} at~$21$ for reversing, and \emo{calm} at~$15$ otherwise. After an initial HTTP/1.0 proxy lost about a quarter of requests through connection resets, the reported implementation uses persistent connections, one retry and fallback to \emo{calm}.

The ten reported campaigns used a second harness host with the same checkpoint, serving image, scaffold and agent image, with eight parallel workers. Each completed $38$ or $39$ challenges.

\paragraph{Log attribution and timing.} The $5{,}419$ decisions are crypto $33\%$, web $27\%$, reversing $21\%$, forensics $10\%$ and misc $9\%$, with none assigned to pwn or other. Attribution requires only one challenge to be active: $1{,}046$ requests meet this condition, spanning $82$ runs with no reversing run. Of these requests, $83\%$ receive their true category's mapped direction (crypto $90\%$, forensics $98\%$, web $99\%$, misc $100\%$, pwn $3\%$, other $0\%$). There are $65$ switches among $964$ consecutive request pairs, across $28$ runs.

Five timed classifications of a $2{,}000$-character state took about $83$\,ms each. The $5{,}318$ timed model requests had median/mean response durations of $3.8/10.5$\,s, compared with $3.0/5.9$\,s for fixed \emo{calm} and $3.1/6.7$\,s for the neutral on the same host. Classification is about $2\%$ of the router's median response. Among attributable runs, mean request intervals, including generation and action, were $3.2$\,s under \emo{calm}, $4.7$ under \emo{serene} and $17.0$ under \emo{desperate} (medians $2$--$3$\,s). The latter accounts for $73\%$ of this time but $36\%$ of requests, chiefly in cryptography.

\end{document}